%% file: Seiler_ImperfectionsRT_2019.tex
\documentclass[3p]{elsarticle}
\usepackage{lineno}
\modulolinenumbers[5]
\journal{Acta Materialia}
\usepackage{graphicx,ulem}
\usepackage{xspace}
\xspaceaddexceptions{)} 
\xspaceaddexceptions{]} 
\usepackage{textcomp}
\usepackage[]{lmodern}
\usepackage{siunitx} 
\usepackage{tabularx}
\usepackage{floatrow}
\usepackage[font={Large,bf},labelformat=simple]{subfig} 
\usepackage{color}
\usepackage{amsmath}
\usepackage{soul} 
\usepackage{booktabs} 
\biboptions{numbers,sort&compress}
\usepackage[plainpages=false]{hyperref} 
\hypersetup{final} 
\usepackage{listings} 
\usepackage{paralist}
\usepackage{epstopdf}
\usepackage{setspace}
\usepackage[export]{adjustbox}
\usepackage{textcomp,fancyhdr,mathtools}
\usepackage[T1]{fontenc}
\newsavebox{\oldell}
\savebox{\oldell}{\ensuremath{\ell}}
\let\temp\rmdefault \usepackage{newpxmath} \let\rmdefault\temp
\renewcommand*{\ell}{\usebox{\oldell}}

\hyphenchar\font=\string"7F 
\newcommand{\simum}[1]{\SI{#1}{\micro m}\xspace}
\newcommand{\figref}[1]{Fig.~\ref{#1}\xspace}

\newcommand{\Secref}[1]{Section~\ref{#1}\xspace}
\renewcommand{\eqref}[1]{Eq.~(\ref{#1})\xspace}

\newcommand{\changes}[1] {\color{black} #1 \color{black}}

\renewcommand{\bar}[1]{\mkern 1.25mu\overline{\mkern-1.25mu#1\mkern-1.25mu}\mkern 1.25mu}

\newcommand{\RMS}{standard deviation\xspace}
\newcommand{\rms}{sd}
\begin{document}

\begin{frontmatter}

  \title{The role of defects in dictating the strength of brittle honeycombs made by rapid prototyping}

  \author[cam]{P.~E.~Seiler}
  \ead{pes34@cam.ac.uk}
  \author[cam]{H.~C.~Tankasala}
  \ead{hct30@cam.ac.uk}
  \author[cam]{N.~A.~Fleck\corref{cor1}}
  \ead{naf1@cam.ac.uk}
  
  \cortext[cor1]{Corresponding author}
  
  \address[cam]{Department of Engineering, University of Cambridge,
    Cambridge CB2 1PZ, United Kingdom}

\begin{abstract}
  Rapid prototyping is an emerging technology for the fast make of
  engineering components. A common technique is to laser cut a
  two-dimensional (2D) part from polymethyl methacrylate (PMMA)
  sheet. However, both manufacturing defects and design defects (such
  as stress raisers) exist in the part, and these degrade its
  strength. In the present study, a combination of experiment and
  finite element analysis is used to determine the sensitivity of the
  tensile strength of PMMA hexagonal lattices to both
  \textit{as-manufactured} and \textit{as-designed} defects. The
  \textit{as-manufactured} defects include variations in strut
  thickness and in Plateau border radius. The knockdown in lattice
  tensile strength is measured for lattice relative density in the
  range of 0.07 to 0.19. A systematic finite element (FE) study is
  performed to assess the explicit role of each type of
  as-manufactured defect on the lattice strength.
  \textit{As-designed} defects such as randomly perturbed joints,
  missing cells, and solid inclusions are introduced within a regular
  hexagonal lattice. The notion of a transition flaw size is used to
  quantify the sensitivity of lattice strength to defect size.
\end{abstract}


\end{frontmatter}

\section{Introduction}
\label{sec:introduction}
Engineering parts made from foams or lattices are increasingly used in
a large variety of engineering applications over a wide range of
length scale, e.\,g. space frames in civil engineering, the cores of
lightweight sandwich structures, and cardiovascular
stents~\cite{Phani2017}. Recent advances in additive manufacturing
methods~\cite{Tumbleston2015,Vyatskikh2018} have enabled the
development of novel two-dimensional (2D) or three-dimensional (3D)
architectured structures with high stiffness and strength relative to
those of foams~\cite{Deshpande2001}. Whilst the compressive behaviour
of lattices and foams has been examined
extensively~\cite{Li2007,Tancogne-Dejean2016,Queheillalt2005,Andrews1999},
only limited studies have been performed on the macroscopic response
and failure criteria under tensile
loading~\cite{Montemayor2016,Olurin2000,Ma2016}. The current study
addresses this gap in the literature and explores the influence of
both \textit{as-manufactured} and \textit{as-designed} defects upon
tensile strength.

\subsection{As-manufactured defects in lattice materials}
\label{sec:role_defects_lattices}
The macroscopic properties of lattice materials are sensitive to the
material properties of individual struts. In turn, the properties of
the struts depend upon the details of the thermal history imposed by
the manufacturing process, such as selective laser melting
(SLM) for example. Consequently, a large scale octet-truss lattice made from the
SLM technique has a lower yield strength than a smaller lattice of the
same relative density~\cite{Tancogne-Dejean2016}. In
the present study, we employ a laser-cutting technique to manufacture
2D hexagonal lattices from amorphous polymethyl methacrylate (PMMA)
sheets. This manufacturing technique is commonly used for the rapid
prototyping of an engineering part, but leads to variability in cell
wall material properties analogous to those of additive techniques
such as selective laser sintering and 3D printing~\cite{Gebhardt2016}.

The sensitivity of modulus and yield strength to imperfections within
a lattice has been studied in systematic fashion, see for example
\cite{Gibson1999, Ronan2016, Symons2008, Schmidt2001, Tankasala2015,
  Tankasala2017, Fleck2007, Romijn2007, Grenestedt2005, Liu2017}. It
is generally accepted that a number of types of imperfection, such as
missing or wavy cell walls degrade the tensile and compressive
properties, although the sensitivity of the dispersion in macroscopic
properties to the statistical distribution of imperfections has
received limited attention. Further, little is known about the
relationship between the macroscopic ductility of a lattice or foam
and its underlying topology. One might expect that the ductility of a
bending-dominated lattice (such as the hexagonal honeycomb) much
exceeds that of the parent solid by the following simple
argument. Assume that the macroscopic response of the lattice can be
idealised by that of a long slender beam (of length $\ell$ and height
$t$), which is built-in at one end and suffers a transverse tip
deflection $\delta$ at the free end, due to a transverse end
load. Then, the bending strain at the built-in end scales as
$(\delta/\ell)$ times a knockdown factor of $t/\ell$. This has the
interpretation that the bending strain as experienced by a material
element of a bending-dominated lattice is much less than the
macroscopic strain (of order $\delta/\ell$). If this were the case,
then the macroscopic ductility of the lattice would much exceed that
of the cell wall material. However, experimental studies indicate that
the ductility of open-celled metallic and polymeric foams are commonly
below that of the solid material. For example, \citet{Ronan2016} have
compiled an overview of the observed tensile ductility of open-cell
foams as a function of relative density (the ratio of the density of
the lattice material to that of the solid). In broad terms, the
measured ductility is 10\% to 50\% that of the parent solid. One
possible source of the knockdown in macroscopic properties is the
presence of geometric imperfections in the form of misaligned struts,
wavy or missing struts, a variation in Plateau border radius and in strut
thickness, misplaced joints, and cell-level inclusions in the form of
filled cells. A major focus of the present study is to explore the
potency of both the as-manufactured and as-designed defects upon the
macroscopic modulus and tensile ductility (or equivalently, the
ultimate tensile strength (UTS)) of an elastic-brittle bending-dominated
lattice as produced by rapid prototyping.

\subsection{Tensile response of a hexagonal lattice}
\label{sec:tensile_response_honeycombs}
A regular two-dimensional (2D) hexagonal lattice is shown
in~\figref{fig:PerfectLatticeSketch}. It comprises struts of length
$\ell$ and in-plane thickness $t$ such that the relative density of
the lattice is given by \cite{Gibson1999}
\begin{equation}
\overline\rho=\frac{2}{\sqrt{3}} \frac{t}{\ell}
\label{eq:relative_density}
\end{equation}
for $t/\ell<0.2$. Under macroscopic uniaxial tensile loading, the
hexagonal lattice is bending-dominated in its small strain elastic
response due to low nodal connectivity of 3, see for
example~\cite{Gibson1999}. However, at a sufficiently high value of
tensile macroscopic strain, the response of the lattice switches from
a compliant, bending-dominated mode to a stiff, stretching-dominated
mode as the cell walls align with the tensile
direction~\cite{Ronan2016,Tankasala2017}. In general, the uniaxial
tensile response of an elastoplastic hexagonal lattice proceeds in 4
stages under an increasing macroscopic strain, as discussed
by~\citet{Ronan2016}: (i) elastic bending of the struts inclined to
the loading axis, (ii) plastic bending, (iii) elastic stretching of
all struts in the lattice as the inclined struts align with the
loading direction, and (iv) plastic stretching. Failure may intervene
during any of these 4 regimes depending upon the active mode of cell
wall failure. \citet{Tankasala2017} have assumed 2 possible criteria
for the failure of a cell wall: (i) when the maximum local strain at
any point in the lattice attains the failure strain of the solid; this
criterion is appropriate for lattices made from ceramics and brittle
alloys, or (ii) when the maximum value of average tensile strain
across any cell wall of the lattice attains the cell wall failure
strain; this criterion is suitable for ductile alloys which undergo
necking, see for example \citet{Onck2004}. The wide range in the
mechanical response of the hexagonal lattice (depending upon the cell
wall material) motivates its choice in the present study. We choose
polymethyl methacrylate (PMMA) for the cell wall material as it is
elastic-brittle at room temperature, but exhibits a strong
visco-plastic characteristic at temperatures close to the glass
transition value. The creep response of PMMA lattices at high
temperature is the subject of a follow-on study.

\subsection{Scope of study}
\label{sec:scope}
The purpose of the current study is to examine experimentally the
tensile response of two-dimensional (2D) elastic-brittle lattice
materials made by rapid prototyping. Lattices were cut from polymethyl
methacrylate (PMMA) sheets and were tested under uniaxial tension at
room temperature. The sensitivity of the macroscopic tensile stress
versus strain curve to as-manufacturing defects, such as variations in
strut thickness and Plateau border radius, was determined for regular
hexagonal lattices. Finite element (FE) calculations were used to
assess the significance of as-manufactured defects on the macroscopic
properties; the precise lattice geometry was determined by
computer-assisted tomography (CT). Irregular lattices were also
created by the introduction of as-designed defects, specifically a
centre crack (due to missing cell walls), solid inclusions in the form
of filled cells, and randomly perturbed joints. The knockdown in
lattice strength due to each of these defects was measured.

\section{Experimental programme}
\label{sec:experimental_procedure}
Specimens were manufactured by the laser-cutting\footnote{\textit{HPC
    Laser Ltd LS6090 Pro} 80 Watt laser cutter; process parameters:
  cutting speed, 60\% power, 55\% corner power.} of cast \SI{5}{mm}
thick PMMA sheets into the following 5 geometries:
\begin{itemize}
\item[(i)] macroscale, dogbone-shaped specimen, as shown
  in \figref{fig:dbTension}, for material characterisation on a large scale;
\item[(ii)] single strut specimen, as shown in
  \figref{fig:singleStrutTension}, for material characterisation on a
  small scale;
\item[(iii)] unit cell specimen of the lattice, comprising a single
  strut embedded in hexagonal cells, see \figref{fig:singleStrut};
\item[(iv)] regular hexagonal lattice, as shown in
  \figref{fig:PerfectLatticeSketch}, to measure the lattice response,
  absent as-designed defects;
\item[(v)] irregular hexagonal lattice containing as-designed defects
  in the form of (a) randomly perturbed joints, (b) missing cell
  walls, or (c) solid inclusions, see \figref{fig:geometries}.
\end{itemize}
  
The strut thickness in specimens of type (ii) to (iv) is
$t=\SI{0.47}{mm}$. For lattice specimens of type~(iv), the relative
density of the lattice $\bar\rho$ is varied from 0.07 to 0.19 by
varying strut length $\ell$ in the range of \SI{3.0}{mm} to
\SI{7.5}{mm}, as demanded by \eqref{eq:relative_density}. The PMMA
material employed in this study has a glass transition
temperature\footnote{The value of $T_\mathrm{g}$ was measured by
  Dynamic Mechanical Analysis (DMA) of a single PMMA cantilever beam
  at an excitation frequency equal to \SI{0.1}{Hz} and a heating rate
  of \SI{5}{\celsius/min}, refer to~\cite{VanLoock2018} for details of
  the test procedure.} $T_\mathrm{g}= \SI{385}{K}$. All specimens were
tested at room temperature $T=\SI{296}{K}=0.77T_\mathrm{g}$.

\subsection{Manufacture of regular hexagonal lattices}
\label{sec:regular_irregular_honeycombs}
A computer-aided drawing (CAD) of the geometry of a regular hexagonal
lattice, as shown in \figref{fig:PerfectLatticeSketch}, was created
using the \textit{OpenSCAD} software. This CAD file provides an input
to the laser cutting machine with sufficient data to define the
translation of the cutting head relative to a fixed position on the
PMMA sheet. The hexagonal lattices were manufactured to a dogbone
shape in order to ensure that failure occurs within the gauge section,
see \figref{fig:PerfectLatticeSketch}. All regular lattice specimens
have a gauge width $W=11\sqrt{3}\ell$ (or 11 cells) and a gauge length
$L=11\ell$ (or 7 cells). \changes{(The same sample size of
  $11\,\times\,7$ cells in the gauge area was used in the follow-on
  study at elevated temperature.)}

\subsection{Test method}
All lattices were tested in uniaxial tension using a screw-driven test
machine at a nominal strain rate of
$\dot\varepsilon=\SI{3.6E-4}{s^{-1}}$. The load $P$ is measured via a
load cell clamped to the stationary platen of the rig while the
extension $u$ of the gauge length is determined by Digital Image
Correlation (DIC), as follows. Prior to the start of the test, the
lattice specimens were coated with a thin layer of white chalk and a
speckle pattern was then generated by spraying black paint in order to
enhance the contrast of the DIC imagery. A single camera of the GOM
ARAMIS 12M system\footnote{maximum resolution: $4096\,\times\,3072$
  pixels, \SI{100}{mm} lens} was used to track facets of size
$20\,\times\,20$ pixels in the vicinity of a predefined node at the
top and bottom of the gauge length. This procedure enabled the
sub-pixel resolution of the nodal displacements.

The uniaxial tensile response of the solid was measured by testing two
geometries: (i) dogbone-shaped solid specimens (of cross-sectional
dimensions $\SI{10}{mm}\,\times\,\SI{5}{mm}$ and gauge length
$L_\mathrm{d}=\SI{80}{mm}$) and (ii) single strut specimens (of mean
strut thickness $\bar{t}=\SI{0.47}{mm}$ and strut length
$L_\mathrm{s}=\SI{10}{mm}$). The extension of the gauge length of the
large dogbone specimens was measured by a non-contact laser
extensometer while that of single strut specimens was measured by
optical tracking of white dots marked on the top and the bottom of the
gauge length along the centre-line of the specimen. The dogbone-shaped
specimens were gripped by wedge grips and the single strut specimens
were pin-loaded.

\section{Material characterisation}
\label{sec:sample_characterisation}
\subsection{Strut geometry}
\label{sec:strut_thickness}
Geometrical features such as the thickness and Plateau border radius
of the individual struts of the laser-cut lattice were measured using
computer-assisted X-ray tomography\footnote{\textit{X-TEK, XT H
    225ST}, voltage: \SI{45}{kV}, current: \SI{200}{mA}, voxel
  size:\simum{18}.}. A probability density function of the strut
thickness $t$ was generated by measuring its value at mid-length on
453 struts of a regular honeycomb lattice of $\bar\rho=0.11$ via image
processing of its CT scan. The strut thickness follows a normal
distribution with a mean value $\bar{t}=\SI{0.47}{mm}$ and a \RMS
$t_{\rm \rms}=\SI{0.09}{mm}$ and this is expressed via the notation
$t=0.47\pm\SI{0.09}{mm}$. We emphasise that the same notation is used
to give the mean value and standard deviation of other quantities such
as peak load, stress and so on. The as-manufactured lattice specimens
also possess a dispersion in the Plateau border radius of each
joint. To quantify this distribution, values of the Plateau border
radius were measured at 300 locations across 10 unit cell specimens;
the smaller size of the unit cell specimens ensured higher spatial
resolution of their CT scans. The Plateau borders have a mean radius
of $\bar{r} = \SI{0.4}{mm}$ and a \RMS $r_{\rm \rms}=\SI{0.1}{mm}$
(not shown).

\subsection{Solid material response} 
\label{sec:solidresponse}
The as-manufactured material properties of solid PMMA were measured
from the response of a large dogbone-shaped specimen and of a single
strut of the lattice; any size effect on the properties is thereby
determined. The nominal stress versus strain response of the solid
dogbone specimens and single strut specimens are compared in
\figref{fig:stressStrain_PMMA_StrutVsSolid}. All specimens respond in
an elastic-brittle manner. The Young's modulus of the solid material,
as measured from the dogbone specimens, is
$E_\mathrm{s}=2.5 \pm \SI{0.2}{GPa}$, and the tensile strength is
$\sigma_\mathrm{fs}=63\pm\SI{2}{MPa}$. In contrast, the Young's
modulus and tensile strength of the solid material, as measured from
the single strut experiments, is $E_\mathrm{s}=2.4\pm\SI{0.3}{GPa}$
and $\sigma_\mathrm{fs}=36\pm\SI{14}{MPa}$.  Whilst the value of
Young's modulus is comparable for the two geometries, the tensile
strength of the single strut is less than that of the larger specimens
by approximately 40\%. This is ascribed to the difference in thermal
histories during manufacture. A significant scatter in the macroscopic
failure strain is also observed for the single strut specimens, such
that the ductility $\varepsilon_\mathrm{fs}=0.015\pm0.007$.

\section{Numerical study}
Quasi-static FE simulations were performed using ABAQUS/Explicit v6.14
to simulate the observed response of the hexagonal lattices under
uniaxial tension. The 2D geometry for each FE model is defined from
the CT scan at the mid-plane of the corresponding as-manufactured 3D
specimen. Uniaxial loading of the lattice is simulated by constraining
all degrees of freedom along the bottom edge of the specimen while the
top edge is subjected to uniform displacement in the $x_{2}$-direction
of the specimen, see \figref{fig:fe_sketch}\!(a). The as-manufactured
lattice of \figref{fig:fe_sketch} contains 453 struts, each of length
$\ell \approx \SI{5}{mm}$. A typical unit cell within the lattice is
shown in the inset of \figref{fig:fe_sketch}\!(a): the struts within
the lattice have variable thickness and variable Plateau border radius
due to the manufacturing process. The FE mesh of the lattice comprises
triangular elements with quadratic shape functions and in-plane strain
(type CPE6M). The elements are of uniform size $\ell_{\rm e}$, chosen
to be such that the thinnest strut in the lattice has at least 7
elements across its thickness, and the stress concentration at the
Plateau borders is adequately captured. A typical FE mesh at a joint
in the lattice is shown in \figref{fig:fe_sketch}\!(a).

The assumed stress $\sigma$ versus strain $\varepsilon$ response of
the cell wall solid is sketched in \figref{fig:fe_sketch}\!(b). A
continuum damage model\footnote{The Johnson-Cook ductile damage model
  is used with parameters to give an elastic-brittle response.} is
employed in the FE calculations to simulate fracture of any element. The
overall response of a material point (i.e. at each integration point of an
element) proceeds in 3 stages as follows:
\begin{itemize}
\item[(i)] The initial undamaged response is linear elastic with
  Young's modulus ${E}_{\rm s}=\SI{2.4}{GPa}$, as taken from the
  measured mean value of single strut specimens, recall
  Section~\ref{sec:solidresponse}. A Poisson's ratio of
  $\nu_{\rm s}=0.3$ is assumed for the cell wall solid.
\item[(ii)] Failure \textit{initiates} at an integration point when
  the maximum tensile stress at that point attains the solid tensile
  strength, $\sigma_{\rm fs}$, as indicated in
  \figref{fig:fe_sketch}\!(b). A deterministic value of
  $\sigma_{\rm fs}=\SI{36}{MPa}$ is assumed for all struts in the
  lattice, as taken from the measured mean tensile strength of single
  strut specimens, recall Section~\ref{sec:solidresponse}.
\item[(iii)] The subsequent \textit{evolution} of damage at a material
  point is specified via a linear softening $\sigma$ versus
  $\varepsilon$ response in accordance with a specified fracture
  energy $\Gamma_{\rm f}$ as defined by
\begin{equation}
  \Gamma_{\rm f}=\ell_{\rm c}\int_{\varepsilon_{\rm fs}}^{\varepsilon_{\rm fs}+\Delta\varepsilon} \sigma \; d\varepsilon
\label{eqn:damagelaw}
\end{equation}
where $\ell_{\rm c}$ is a characteristic length associated with the
integration point of the element: $\ell_{\rm c}$ equals $\ell_{\rm e}/2$ for
quadratic triangular finite elements, $\varepsilon_{\rm fs}$ equals
$\sigma_{\rm fs}/E_{\rm s}$, and $\Delta\varepsilon$ is the strain
increment over the softening portion of the response, see
\figref{fig:fe_sketch}\!(b). A triangular damage evolution law is
assumed in the present study such that
$\Delta\varepsilon=2\Gamma_{\rm f}/\sigma_{\rm fs}\ell_{\rm c}$. Note
that there is choice in the precise value of $\Gamma_{\rm f}$; the FE
simulations reported here assume $\Gamma_{\rm f}=0.1 \ \rm kJ/m^{2}$
following \cite{Choi1993}. It is noted that the specification of damage
evolution via \eqref{eqn:damagelaw} in terms of the size of the finite
element alleviates the problem of mesh dependence of the solution;
refer to \cite{Oliver1989} for details.
\end{itemize}

\section{Tensile response of as-manufactured lattice specimens:
  prediction versus experiment}
\label{sec:experimental_results}
The measured tensile response of the two geometries of specimens, the
unit cell specimen of \figref{fig:singleStrut} and the lattice
specimen of \figref{fig:PerfectLatticeSketch}, are discussed in
turn. Comparisons with the FE predictions are made alongside.

\subsection{Unit cell specimen}
\label{sec:single_strut_specimen}
Consider the unit cell specimen of \figref{fig:singleStrut}. We
characterise the structural response of this specimen in terms of its
load $P$ versus overall displacement $u$, and include the
as-manufactured defects in the FE predictions of the structural
response. In the experiments, two support struts, spanning the height
of the specimen on either side of the central strut, protect the
specimen prior to the test; these are cut at their mid-length
immediately prior to the test. A total of 10 specimens were tested
under uniaxial tension along the axis of the central strut, see
\figref{fig:singleStrut}. All struts are of length $\ell=\SI{5}{mm}$
and the mean thickness of a strut (across the 10 specimens) is
$\bar{t}=\SI{0.47}{mm}$. The measured load $P$ versus displacement $u$
response of 3 representative unit cell specimens is shown in
\figref{fig:fe_pu_singlestrut}. In each case, the response is shown
until failure of the first strut. The peak load is
$P_\mathrm{max}=28.6 \pm \SI{5.4}{N}$ from the measured values of 10
specimens. Failure of all specimens (except one) occurs in an inclined
strut adjacent to the central vertical strut and close to the joint.

FE predictions of the tensile response of 3 unit cell specimens are
included in \figref{fig:fe_pu_singlestrut}; the 2D geometry of each
specimen is determined from a CT scan of the mid-plane of the
corresponding as-manufactured specimen. A constant value of strut
tensile strength, $\sigma_{\rm fs}=\SI{36}{MPa}$, is assumed for all
struts in the specimen, as taken from the mean value of the measured
tensile strength of the single strut specimens, recall
Section~\ref{sec:solidresponse}. We find from
\figref{fig:fe_pu_singlestrut} that the predicted $P$ versus $u$
response is consistent with the measured response for specimens 1 and
2, indicating that \changes{the dispersion in geometry
  (i.\,e. variation in strut thickness and Plateau border radius)
  largely explains the observed response of the unit cell
  specimens. The strut tensile strength is likely to be uniform in the
  unit cell specimens for the following reason. The variance in the
  strut strength is a consequence of thermal history associated with
  strut manufacture. The time interval between laser-cutting of the
  first and the last struts of the specimen is only 20 seconds for the
  unit cell specimens and consequently, there is good repeatability in
  strength from strut to strut.}  Agreement is less satisfactory for
specimen 3, for reasons unknown. The location of first strut failure,
as seen in the 3 representative specimens, is sketched in
\figref{fig:fe_failure_singlestrut}: the circles denote the location of
first strut failure in experiments while crosses are the FE
predictions. A single strut in the FE simulation is considered to have
failed when a row of finite elements at any strut cross-section become
stress-free, following \eqref{eqn:damagelaw}. In general, the critical
strut that fails first is an inclined strut adjacent to the central
vertical strut, in both experiments and FE.

\changes{Both experiments and FE reveal that failure occurs within the
  inclined struts at a location near the joints. Thus, the FE
  indicates that the local stress in the inclined struts exceeds that
  of the central strut.}

\subsection{Lattice specimens}
\label{sec:lattice_specimens}
The measured macroscopic nominal stress $\sigma^{\infty}$ versus
nominal strain $\varepsilon^\infty$ response of one lattice specimen
(labelled \textit{Test~I}) of relative density $\bar{\rho}=0.11$ is
shown in \figref{fig:fe_exp_stressstrain_47}, with $\sigma^{\infty}$
and $\varepsilon^\infty$ defined in terms of the reaction force $P$ on
the top edge and the extension $u$ of the gauge length as
$\sigma^{\infty}=P/(WB_0)$ and $\varepsilon^\infty=u/L$,
respectively. The gauge dimensions $(W,L)$ are indicated in
\figref{fig:fe_sketch}; $W=\SI{95.3}{mm}$, $L=\SI{55}{mm}$, and
$B_0=\SI{5}{mm}$ for the specimen of
\figref{fig:fe_exp_stressstrain_47}. The FE prediction of the
macroscopic response is included in
\figref{fig:fe_exp_stressstrain_47}. As before, the geometry of the FE
model is determined from a CT scan of the mid-plane of the
as-manufactured specimen. A constant tensile strength of
$\sigma_{\rm fs}=\SI{36}{MPa}$ is assumed for all struts in the
lattice, as taken from the mean value of the measured tensile strength
of the single strut specimens in Section~\ref{sec:solidresponse}. The
sequence of strut failure is shown in
\figref{fig:fe_exp_stressstrain_47} for both measurement and
prediction: the number indicates the position of struts in the failure
sequence with their actual location marked in
\figref{fig:fe_failure_lattice_caseE}. We find from
\figref{fig:fe_exp_stressstrain_47} and
\figref{fig:fe_failure_lattice_caseE} that an initial drop in load
accompanies failure of a strut located at the edge of the
specimen. Subsequent load drops correspond to the simultaneous failure
of multiple struts in the vicinity of previously failed
struts. Eventually, a dominant macroscopic crack propagates across the
specimen leading to catastrophic fracture. The FE predictions for the
initial stiffness and the load corresponding to first strut failure
are in close agreement with the corresponding values in the
experiment, despite the mismatch in the precise location of strut
failure, see \figref{fig:fe_exp_stressstrain_47} and
\figref{fig:fe_failure_lattice_caseE}.

Additional tension tests were performed on 4 lattice specimens (also
of $\bar\rho=0.11$). The measured and predicted
values of the macroscopic stress corresponding to
the first strut failure, and the peak stress
are shown in \figref{fig:fe_exp_stresses_mod} for all 5 lattice
specimens. In all cases, failure of the lattice occurs in a correlated
manner (i.\,e. via the formation of a single horizontal crack
originating from one edge of the specimen) subsequent to first strut
failure. FE simulations of the 4 specimens also predict this pattern
of failure. Visual examination of the failed specimens further
revealed that the struts always fail close to a joint, and that the
thinner struts fail first: the mean thickness of failed struts is
about 10\% lower than the overall mean strut thickness of
\SI{0.47}{mm}.

\changes{\subsection{Effect of relative density on the macroscopic
    properties}}
\label{sec:effect_specimen_size}
It is instructive to evaluate the accuracy of the scaling laws for the
Young's modulus and tensile strength of a lattice material of infinite
extent with those of the laser-cut finite specimens of the present
study. Consider a regular hexagonal lattice comprising struts of
length $\ell$ and thickness $t$, and made from an elastic-brittle
solid of Young's modulus $E_\mathrm{s}$ and tensile strength
$\sigma_\mathrm{fs}$. The macroscopic in-plane modulus $E$ and tensile
strength ${\sigma^\infty_\mathrm{f}}$ of an infinite hexagonal lattice
are given by \cite{Gibson1999}
\begin{equation}
  \label{eq:youngs_modulus}
  {E}=\dfrac{3}{2}\bar\rho^3 E_\mathrm{s} \quad \textrm{and} \quad {\sigma^\infty_\mathrm{f}}=\dfrac{1}{3}\bar{\rho}^2  {\sigma_\mathrm{fs}}
\end{equation}
Lattice specimens of geometry identical to those as shown in
\figref{fig:PerfectLatticeSketch} were laser-cut from PMMA sheets, in
order to verify the accuracy of \eqref{eq:youngs_modulus} as a
function of the relative density of the lattice. In each case, the
strut thickness $t$ was held constant such that
$\bar{t}=\SI{0.47}{mm}$ for all struts, and the strut length $\ell$
was varied in accordance with \eqref{eq:relative_density} to generate
lattices of $\bar\rho=0.07$ and $\bar\rho=0.19$.

The measured values of macroscopic modulus $E$ and tensile strength
$\sigma^\infty_\mathrm{f}$ of the lattice specimens are plotted in
\figref{fig:scaling_laws} as a function of $\bar\rho$ for 5 specimens
per value of $\bar\rho$, upon \changes{normalising $E$ and
  $\sigma_\mathrm{f}^\infty$ by } $\bar{E}_\mathrm{s}=\SI{2.5}{GPa}$
and $\bar\sigma_\mathrm{fs}=\SI{36}{MPa}$\changes{, respectively.} The
predictions of \eqref{eq:youngs_modulus} are included in
\figref{fig:scaling_laws} for comparison. It is clear that both $E$
and $\sigma^\infty_\mathrm{f}$ scale with $\bar\rho$ according to
\eqref{eq:youngs_modulus}. Deviation of the measured values of $E$ and
$\sigma^\infty_\mathrm{f}$ from \eqref{eq:youngs_modulus}, and the
scatter in data for each value of $\bar\rho$ are attributed to the
as-manufactured defects such as the variation in strut thickness $t$
and variation of Plateau border $r$ in addition to the finite size of
the specimen. \changes{The measured values of the macroscopic Young's
  modulus and the macroscopic tensile strength as obtained from
  dogbone specimens, with a gauge size of 13 cells (in height) and 7
  cells (in width), are included in \figref{fig:scaling_laws} along
  with the case of 11\,$\times$\,7 cells for lattices of
  $\bar{\rho}=0.11$. The observed insensitivity of the macroscopic
  elastic response with choice of specimen geometry, as seen in
  \figref{fig:scaling_laws}, is consistent with the fact that the
  elastic boundary layer in a hexagonal lattice spans only one cell
  size.}

\changes{It remains to determine by FE analysis the explicit role of
  each type of as-manufactured defect in reducing the macroscopic
  tensile strength. We achieve this by the sequential introduction of
  two classes of geometric imperfections, (i) a dispersion in strut
  thickness and (ii) a dispersion in Plateau border. The details are
  as follows.}

\section{Effect of as-manufactured defects upon the macroscopic
  tensile strength: prediction versus experiment}
The role of geometric imperfections, such as a dispersion in strut
thickness and in Plateau border radius, is now predicted by
considering the following lattice geometries, with the salient details
of each geometry as summarised in Table~\ref{table:knockdown}.
\begin{itemize}
\item[(i)]\textit{ Case~A }is the reference case of a geometrically
  perfect lattice with a deterministic, uniform value of strut
  thickness $t=\bar{t}=\SI{0.47}{mm}$ and of Plateau border radius
  $r=\bar{r}=\SI{0.4}{mm}$.
\item[(ii)] \textit{Case~B} is an imperfect lattice, with a normal
  distribution of the Plateau border radius such that the mean radius
  in the specimen is $\bar{r}=\SI{0.4}{mm}$ and standard deviation
  $r_{\rm \rms}/\bar{r}=0.25$, based on measurements of the
  as-manufactured specimens, as discussed previously in
  Section~\ref{sec:strut_thickness}. All struts of this geometry have
  a constant value of strut thickness $t=\bar{t}=\SI{0.47}{mm}$, as
  for case~A.
\item[(iii)] \textit{Case~C} is an imperfect lattice, with a normal
  distribution of the strut thickness such that
  $\bar{t}=\SI{0.47}{mm}$ and $t_{\rm \rms}/\bar{t}=0.19$, based on
  measurements of the as-manufactured specimens of
  Section~\ref{sec:strut_thickness}. All struts of this geometry have
  a constant value of Plateau border radius $r=\bar{r}=\SI{0.4}{mm}$,
  as for case~A.
\item[(iv)] \textit{Case~D} is an imperfect lattice, with a normal
  distribution of \textit{both} the strut thickness and the Plateau
  border radius such that $\bar{t}=\SI{0.47}{mm}$,
  $\bar{r}=\SI{0.4}{mm}$, $t_{\rm \rms}/\bar{t}=0.19$, and
  $r_{\rm \rms}/\bar{r}=0.25$. All struts have a uniform thickness
  along their length. 
\item[(iv)] \textit{Case~E} is a set of nominally identical
  as-manufactured geometries. Each is defined from a CT scan of the
  mid-plane section of the corresponding as-manufactured lattice
  specimen, as previously discussed in
  Section~\ref{sec:lattice_specimens}. We emphasise that each
  as-manufactured geometry contains imperfections resulting from the
  manufacturing process (such as wavy or misaligned struts, and
  non-uniform thickness along the strut length). The distribution of
  strut thickness and Plateau border radius for this set of
  as-manufactured geometries has already been characterised as
  $\bar{t}=\SI{0.47}{mm}$, $\bar{r}=\SI{0.4}{mm}$,
  $t_{\rm \rms}/\bar{t}=0.19$, and $r_{\rm \rms}/\bar{r}=0.25$ in
  Section~\ref{sec:strut_thickness}.
\end{itemize}
In all cases, the strut length equals $\ell=\SI{5}{mm}$ such that
$\bar\rho=0.11$, and a constant value of strut tensile strength is
assumed such that $\sigma_{\rm fs}=\SI{36}{MPa}$. The FE predictions
of the lattice tensile strength $\sigma^{\infty}_{\rm f}$ (based on
failure of the first strut) are listed in Table~\ref{table:knockdown}
for cases~A to D; data are presented in terms of the mean strength
$\bar\sigma^{\infty}_{\rm f}$ and standard deviation
${\sigma}^{\infty}_{\rm f,sd}$ from 10 realisations. The values in
each case are normalised by the mean measured strength of the lattice,
$\bar\sigma^{\infty}_{\rm Exp}$.  The predicted values of tensile
strength for 5 realisations of case~E are included in
Table~\ref{table:knockdown} for comparison. It is clear from
Table~\ref{table:knockdown} that the introduction of geometric
imperfection reduces the mean value of the macroscopic tensile
strength: a dispersion in thickness alone (case~C) reduces the tensile
strength by up to $20\%$ compared to that of the perfect lattice
strength (case~A). An additional dispersion in Plateau border radius
(case~D) only slightly reduces the mean value of the tensile strength,
but it increases the scatter in strength, compare cases~B and D in
Table~\ref{table:knockdown}. We conclude from
Table~\ref{table:knockdown} that dispersion in strut thickness is the
primary source of the reduction in the mean tensile strength of
laser-cut PMMA lattices.

\subsection{Dispersion in strut tensile strength}
\label{sec:influence_dispersion}
It is evident from Table~\ref{table:knockdown} that the scatter in
${\sigma}^{\infty}_{\rm f}$ for the as-manufactured lattices (case E,
Exp.)  is under-predicted in the FE simulations (case E). Assuming
that the geometrical as-manufactured imperfections are accurately
captured by the FE simulations, a probable source of the observed
scatter in ${\sigma}^{\infty}_{\rm f}$ for case~E experiments is the
variation in tensile strength of the strut material in the lattice,
which is a consequence of the thermal history imposed by the
laser-cutting process\footnote{Note that the unit cell specimens
  differ from the lattice specimens in terms of the contrasting
  temperature history of the struts during manufacture: the time
  interval between laser-cutting of the first and the last struts of
  the specimen is about 20 seconds for the unit cell specimens and
  about 12 minutes for the lattice specimens. The strut tensile
  strength is thus (relatively) uniform in one unit cell
  specimens.}. To assess the role of strength dispersion, FE
simulations were performed on 10 realisations of a geometrically
perfect lattice (case~A) and for an assumed distribution of strut
tensile strength of mean value $\bar{\sigma}_{\rm
  fs}=\SI{36}{MPa}$. Two choices of \RMS of the distribution of strut
strength were assumed:
$\sigma_{\rm fs,\rms}/\bar{\sigma}_{\rm fs}=0.1$ and $0.2$. The normal
distribution curve of strut strength in each case is truncated such
that the strut strength in the lattice varies from \SI{22}{MPa} to
\SI{50}{MPa}, based on the observed minimum and maximum strut tensile
strength, respectively, from the single strut tests (recall
\Secref{sec:solidresponse}). The resulting probability density
functions of strut strength are shown in
\figref{fig:fe_pdf_strutdispersion} for
$\sigma_{\rm fs,\rms}/\bar{\sigma}_{\rm fs}=0.1$ and $0.2$.

The FE predictions of the normalised macroscopic tensile strength for
10 realisations are plotted in \figref{fig:fe_sf_strutdispersion} for
the two selected values of
$\sigma_{\rm fs,\rms}/\bar{\sigma}_{\rm fs}$. The dispersion in the
strut tensile strength leads to a significant reduction in the mean
value of the tensile strength by up to $20\%$ (for
$\sigma_{\rm fs,\rms}/\bar{\sigma}_{\rm fs}=0.2$) but only mildly
increases the scatter. We conclude from Table~\ref{table:knockdown}
and \figref{fig:fe_sf_strutdispersion} that the dispersion in strut
thickness and in strut tensile strength are each potent sources of
knockdown in the macroscopic strength of the as-manufactured PMMA
lattices.

\section{The effect of \textit{as-designed} defects on macroscopic
  tensile strength}
\label{sec:effect_as_designed}
Three types of macroscopic defect were introduced within the regular
lattice by design: (i) misplaced joints, (ii) cells filled with solid
inclusions, and (iii) missing cell walls. The resulting
as-manufactured specimens contain geometric imperfections at the cell
wall level (variation in strut thickness and Plateau border radius) in
addition to one of the three macroscopic defects, see
\figref{fig:geometries}.  The macroscopic tensile strength of the
as-manufactured specimens was measured experimentally and then
compared with that of the as-manufactured
topologies designed without macroscopic defects. The sensitivity of
measured tensile strength to the presence of as-designed defects was
thereby assessed.

\subsection{Randomly perturbed joints}
\label{sec:perturbations_nodal}
Imperfect hexagonal lattices were manufactured by first generating a
CAD file of a lattice with randomly perturbed joints. To achieve this,
the joints of a regular hexagonal lattice (of $\bar\rho=0.11$) were
repositioned randomly within a circular disc of radius $R$, following
the procedure as used by \citet{Romijn2007}. The degree of
imperfection was varied by selecting values of $R/\ell$ between $0$ (
regular lattice) and $0.5$ (extremely imperfect lattice). A typical
realisation of the as-manufactured lattice, for the choice
$R/\ell=0.5$, is shown in \figref{fig:Model_Seed1_0_Radius0_5}; only
those joints which lie within the gauge section are misplaced. The
random misplacement of the joints reduces the average strut length
such that the relative density of the lattice increases by a factor of
1.0025 for $R/\ell=0.1$ and by a factor of 1.0625 for $R/\ell=0.5$, as
previously noted by \citet{Romijn2007}; this minor change in
$\bar\rho$ is ignored in the current study.

The sensitivity to random perturbation of joints is measured for 3
macroscopic properties of the imperfect lattice: (i) Young's modulus
$E$, (ii) Poisson's ratio $\nu$, and (iii) tensile strength
$\sigma^\infty_\mathrm{f}$. These observed sensitivities are plotted
in \figref{fig:misplaced_nodes} as a function of the degree of
imperfection $R/\ell$, for the choice of $\bar\rho=0.11$. The ordinate
in each case (except for $\nu$) is normalised by its corresponding
mean value as measured for the regular lattice ($R/\ell=0$). Results
are shown for 3 realisations of imperfect lattice for each choice of
$R/\ell=0, 0.3,$ and $0.5$. The Young's modulus $E$ of the imperfect
lattice increases with increasing $R/\ell$ as previously noted by
\citet{Ronan2016}: the mean value of $E$ increases by 45\% for
$R/\ell=0.3$, and by 50\% for $R/\ell=0.5$, compared to its value at
$R/\ell=0$. The observed reduction of $\nu$ with increasing $R/\ell$
supports the prediction of \citet{Romijn2007} (not shown). The
macroscopic tensile strength $\sigma^\infty_\mathrm{f}$ is almost
insensitive to the value of $R/\ell$.

The sequence of strut failure is shown in
\figref{fig:hexlattice_cut_26_failure} for a representative specimen
of $R/\ell=0.5$. For each of the imperfect lattices considered in this
study, it was observed that catastrophic failure occurs by the
formation of a single crack originating from the centre of the
specimen, and not at the edge as observed in the regular lattices of
\figref{fig:fe_failure_lattice_caseE}. Subsequent to initial strut
failure of the imperfect lattice, neighbouring struts failed. In all
cases, struts failed close to the joints, consistent with
bending-dominated failure.

\subsection{Missing cell walls}
\label{sec:missing_cell_walls}
Centre-cracked lattice specimens were manufactured with a row of
missing cell walls at the centre of a regular lattice, see
\figref{fig:RemovedCellWallsSketch}. The initial crack is of
semi-length $a_0=\sqrt{3} n_\mathrm{b} \ell /2$ where $n_{\rm b}$ is
the number of broken cell walls. The influence of crack length on the
tensile strength of the lattice was explored by varying $n_{\rm b}$
between 0 and 3; three realisations of the lattice were generated for
each value of $n_\mathrm{b}$.  The measured values of the notch tensile
strength $\sigma^\infty_\mathrm{f}$ are plotted in
\figref{fig:removed_cellswalls_stress} as a function of the crack
length $a_0/\ell$, for the choice $\bar\rho=0.11$. A significant drop
in $\sigma^\infty_\mathrm{f}$ is observed when 1 or more struts are
broken.

The sensitivity of tensile strength of the lattice to missing cell
walls can be rationalised in terms of its transition flaw size
$a_{\rm T}$: this is the semi-length of the crack beyond which
fracture of the lattice is given by a fracture mechanics
assessment. The value of $a_{\rm T}$ for an elastic-brittle lattice
depends upon the fracture toughness of the lattice $K_{\rm IC}$ and
the tensile strength of an intact, crack-free lattice
$\sigma^\infty_\mathrm{f}(a_0=0)$ as \cite{Tankasala2017}
\begin{equation}
  \label{eq:trans_flaw_size}
  a_{\rm T}=\frac{1}{\pi}\left[ \frac{K_{\rm IC}}{\sigma^\infty_{\rm f}(a_0=0)} \right]^2
\end{equation}
where $K_{\rm IC}$ is expressed in terms of the tensile strength of
the cell wall solid $\sigma_{\rm fs}$, relative density of the lattice
$\bar\rho$ and strut length $\ell$ according to \cite{Romijn2007}
\begin{equation}
  \label{eq:kic}
  K_{\rm IC}=0.9\bar\rho^2 \sigma_{\rm fs} \sqrt{\ell}
\end{equation} 
Upon substituting \eqref{eq:kic} into \eqref{eq:trans_flaw_size} and
taking $\sigma^\infty_{\rm f}(a_0=0)$ to be the measured mean strength
of crack-free lattice specimens (of $\bar\rho=0.11$),
$\sigma^\infty_{\rm f}(a_0=0)=\SI{0.2}{MPa}$ (see
\figref{fig:relativeDensity_stressFailure_perfectHex}), we find that
$a_{\rm T}=1.4\ell$ for a hexagonal lattice made from solid PMMA of
tensile strength $\sigma_{\rm fs}=\SI{36}{MPa}$. Predictions for the
tensile strength,
$\sigma^\infty_{\rm f}(a_0)=K_{\rm IC}/\left(Y\sqrt{\pi a_0}\right)$,
are included in \figref{fig:removed_cellswalls_stress} and they are in
good agreement with the measured values. Note that these predictions
include the geometric $K-$calibration factor $Y(a_0/W, L/W)$ from
~\citet{Liu1996} to account for the effect of the finite geometry of
the specimen. We find from \figref{fig:removed_cellswalls_stress} that
the transition from strength-controlled fracture to $K-$controlled
fracture occurs at $a_0 \approx a_{\rm T} =1.4 \ell$. For all values
of $a_0$ explored in this study, a critical strut fails adjacent to
the pre-crack followed by a series of strut failures in its immediate
vicinity, see \figref{fig:hexlattice_cut_32_failure} for the case of
$a_0/\ell=2.6$, in support of this observation.

\subsection{Solid inclusions}
\label{sec:solid_inclusions}
Hexagonal lattices containing a solid inclusion were generated by the
laser-cutting of PMMA sheets, with a number of intact filled cells at
the centre of the specimen, recall
\figref{fig:filled_cells_stress}. The semi-length of the inclusion is
$a_0=\sqrt{3}n_{\rm c}\ell/2$ where $n_{\rm c}$ is the number of
filled cells. Three realisations of the lattice were generated for
each value of $n_{\rm c}$ between 0 and 3. The measured values of
macroscopic tensile strength $\sigma^\infty_\mathrm{f}$ are plotted in
\figref{fig:filled_cells_stress} as a function of the inclusion size
$a_0/\ell$: $\sigma^\infty_\mathrm{f}$ is almost insensitive to the
presence of solid inclusions. The location of the failure is remote
from the inclusion and it is from the edge of the specimen consistent
with the predictions of \citet{Chen2001a}, see
\figref{fig:hexlattice_cut_21_failure} for the case of
$a_0/\ell=2.6$.

\section{Concluding remarks}
\label{sec:conclusin}
The present study explores the relative sensitivity of tensile
strength of a lattice to as-manufactured defects and as-designed
defects. It is found that both classes of defects are significant for
an elastic-brittle lattice that has been manufactured by a rapid
prototyping route. FE analysis provides insight into the relative
potency of defects in controlling both the mean tensile strength and
the dispersion in tensile strength of PMMA lattices.

Hexagonal lattices were generated by laser-cutting PMMA sheets, and
were tested under uniaxial tension at room temperature; at this
temperature, PMMA behaves in an elastic-brittle manner. It is found
from both experiments and FE predictions that the uniaxial tensile
response of the laser-cut PMMA lattices is linear elastic until a
critical strut fails at the edge of the finite specimen (but within
the gauge dimensions), recall
\figref{fig:fe_failure_lattice_caseE}. In turn, a series of struts
adjacent to this failed strut fail in sequence within a relatively
small ($<0.5\%$) increment of macroscopic strain, thereby forming a
single dominant crack at the edge of the specimen. Such behaviour is
consistent with the small transition flaw size
($a_{\rm T} \approx \ell$) for the elastic-brittle hexagonal lattice,
as previously discussed by \citet{Tankasala2017}. In all lattice
specimens tested in this study, failure occurs first in an inclined
strut close to the joint, and all failed struts exist within the gauge
dimensions of the specimen. The measured values of macroscopic Young's
modulus $E$ and macroscopic tensile strength $\sigma^{\infty}_{\rm f}$
scale with the lattice relative density $\bar{\rho}$ in accordance
with the analytical predictions of \citet{Gibson1999} based on a
point-wise stress-based failure criterion for lattice tensile
strength.

Three-dimensional computed tomography analysis of the laser-cut
specimens revealed imperfections in the lattice geometry in the form
of a dispersion in strut thickness and in Plateau border radius. The
influence of these as-manufactured defects was explored by combined
experimental and numerical studies. In general, the presence of these
defects leads to a reduction in the macroscopic tensile strength of
the lattices; a dispersion in strut thickness leads to the largest
knockdown in strength. Catastrophic fracture of the specimen occurs by
the formation of a single crack emanating from the edge of the
specimen and a series of struts fail at an almost-constant value of
load. The dispersion in strut thickness is associated with a
dispersion in strut tensile strength due to the thermal history
imposed on each strut by the laser-based manufacturing
process. Additional FE simulations were performed to explore the
sensitivity of the macroscopic tensile strength to the dispersion in
strut tensile strength. A significant knockdown in macroscopic
strength is predicted by the introduction of a minor dispersion in the
strut tensile strength.

Imperfections were also introduced in the laser-cut lattices
\textit{by design}. Three kinds of imperfections were explored
experimentally: randomly misplaced joints, a row of missing cell walls
to create a notch, and a row of cells filled by solid inclusions. The
following conclusions can be drawn for each type of defect:
\vspace{-0.2cm}
\begin{itemize}
\item[(i)] \textit{Imperfections in the form of randomly perturbed
    joints.} The macroscopic modulus of the lattice increases by
  $40\%$ as $R/\ell$ increases from 0 to 0.5. This finding is in
  agreement with the numerical prediction of \citet{Chen1999} who have
  reported a similar increase in modulus of an elastic-brittle Voronoi
  honeycomb compared to that of the regular hexagonal honeycomb. In
  contrast, the macroscopic tensile strength of the lattice is almost
  insensitive to the perturbation of nodal position and this is
  ascribed to the fact that the primary mode of deformation within a
  failed strut is bending for all values of $R/\ell$ between 0 and
  0.5.
\item[(ii)] \textit{Imperfections in the form of missing cell walls.}
  The notch tensile strength of the centre-cracked lattice ranges from
  the notch-insensitive limit
  $\sigma^{\infty}_{\rm f}(a_0)=\sigma^{\infty}_{\rm f}(0)$ to the
  value based on fracture toughness,
  $\sigma^{\infty}_{\rm f}(a_0)=K_{\rm IC}/(Y\sqrt{\pi a_0})$
  for increasing semi-crack length $a_0$ from $0$ to $3\ell$. The
  transition flaw size is $a_{\rm T} \approx \ell$, consistent with
  the numerical predictions of \citet{Tankasala2017} for an
  elastic-brittle hexagonal lattice. In each case, failure of the
  lattice occurs in a crack-like manner by the successive failure of a
  series of inclined struts ahead of the initial crack tip.
\item[(iii)] \textit{Imperfections in the form of solid inclusions.}
  The lattice tensile strength is insensitive to the presence of
  filled cells; no distinct knockdown in strength is observed for
  inclusions of size $0$ and $3\ell$. Further, the location of
  first strut failure is remote from the inclusions and at the edge of
  the specimen.
 \end{itemize}
 In summary, the macroscopic tensile strength of the hexagonal PMMA
 lattice is most sensitive to imperfections in the form of broken cell
 walls owing to its low transition flaw size. In contrast, only a mild
 sensitivity of the tensile strength is observed for the cases of
 misplaced joints and cells filled with solid inclusions.

\section*{Acknowledgements}
\noindent The authors gratefully acknowledge the financial support
from the European Research Council (ERC) under the European Union's
Horizon 2020 research and innovation program, grant GA669764,
MULTILAT.

\bibliographystyle{elsarticle-num-names}
\bibliography{LiteratureCambridge}

\newpage \input{Tables}

\newpage \input{Figures}

\end{document}

%% file: Tables.tex
\section*{Tables}

\begin{table}[h]
  \begin{tabular}{ l | c | c | c | c }
    case & $t_\mathrm{\rms}/\bar{t}$ & $r_\mathrm{\rms}/\bar{r}$ & $\bar{\sigma}^\infty_\mathrm{f}/\bar{\sigma}^{\infty}_\mathrm{Exp}$ & ${\sigma}^\infty_\mathrm{f,sd}/\bar{\sigma}^{\infty}_\mathrm{Exp}$ \\
    \hline
    \textbf{A} & 0 & 0 & 1.42 & 0\\
    \textbf{B} & 0 & 0.25 & 1.29 & 0.02\\
    \textbf{C} & 0.19 & 0 & 1.18 & 0.05\\
    \textbf{D} & 0.19 & 0.25 & 1.11 & 0.06\\
    \textbf{E} & 0.19 & 0.25 & 1.09 & 0.09\\
    \textbf{E (Exp.)} & 0.19 & 0.25 & 1 & 0.17\\
  \end{tabular}
  \caption{Effect of as-manufactured defects on the tensile strength
    of lattice specimens. In all cases, $\bar{t}=\SI{0.47}{mm}$ and
    $\bar{r}=\SI{0.4}{mm}$.}
  \label{table:knockdown}
\end{table}


%% file: Figures.tex
\section*{Figures}

\floatsetup[figure]{subcapbesideposition=top}

\begin{figure}[htbp]
  \sidesubfloat[]{\includegraphics[width=0.65\textwidth]{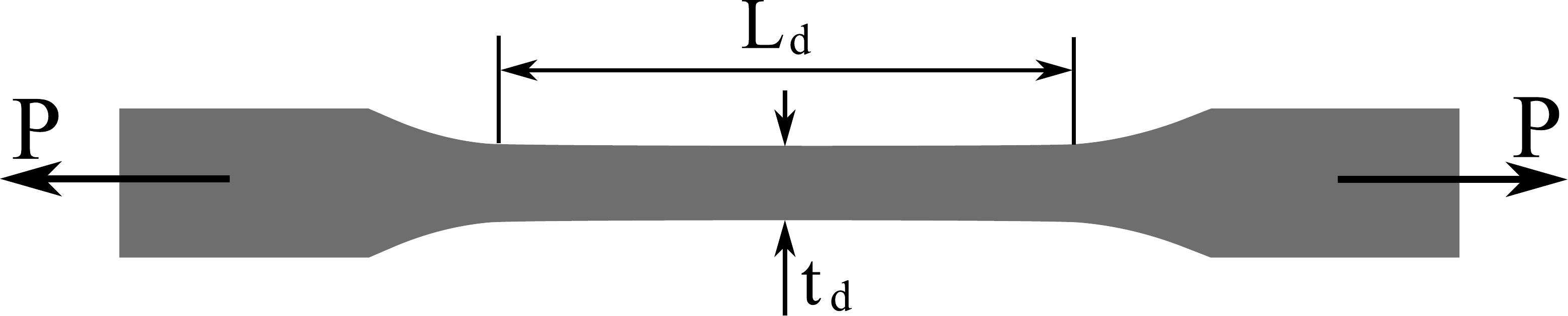}\label{fig:dbTension}}\\[\baselineskip]
  \sidesubfloat[]{\includegraphics[width=0.4\textwidth]{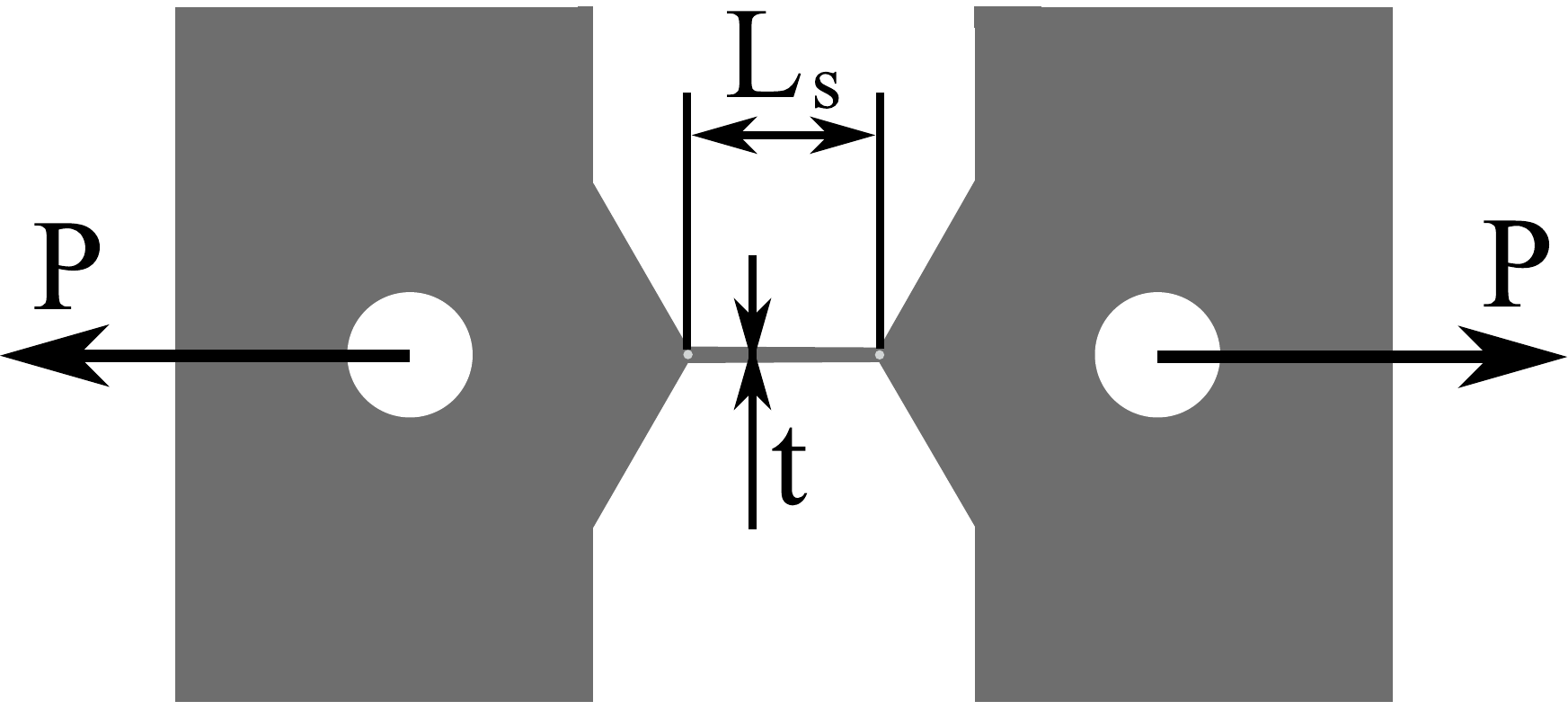}\label{fig:singleStrutTension}}
  \sidesubfloat[]{\includegraphics[width=0.405\textwidth]{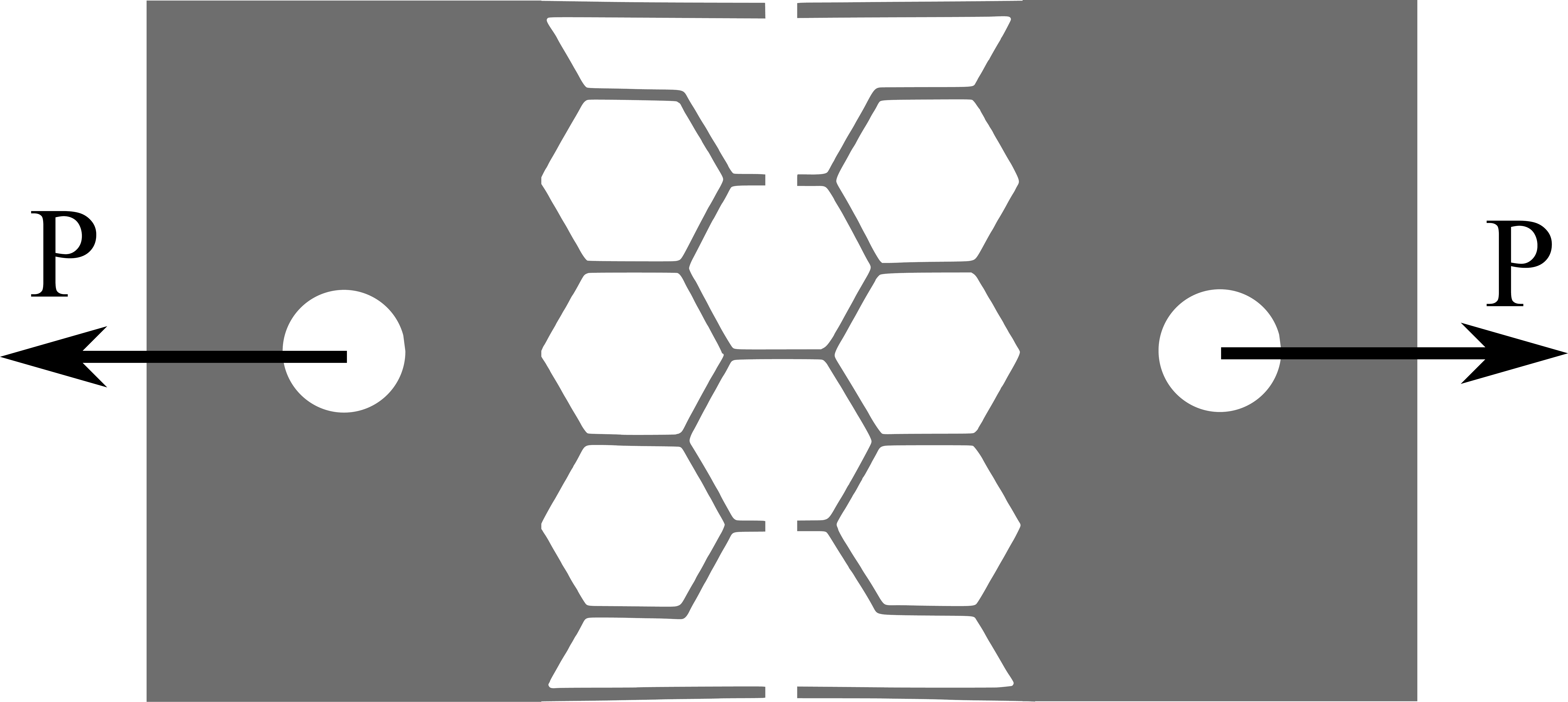}\label{fig:singleStrut}}\\[\baselineskip]
  \sidesubfloat[]{\includegraphics[width=0.55\textwidth]{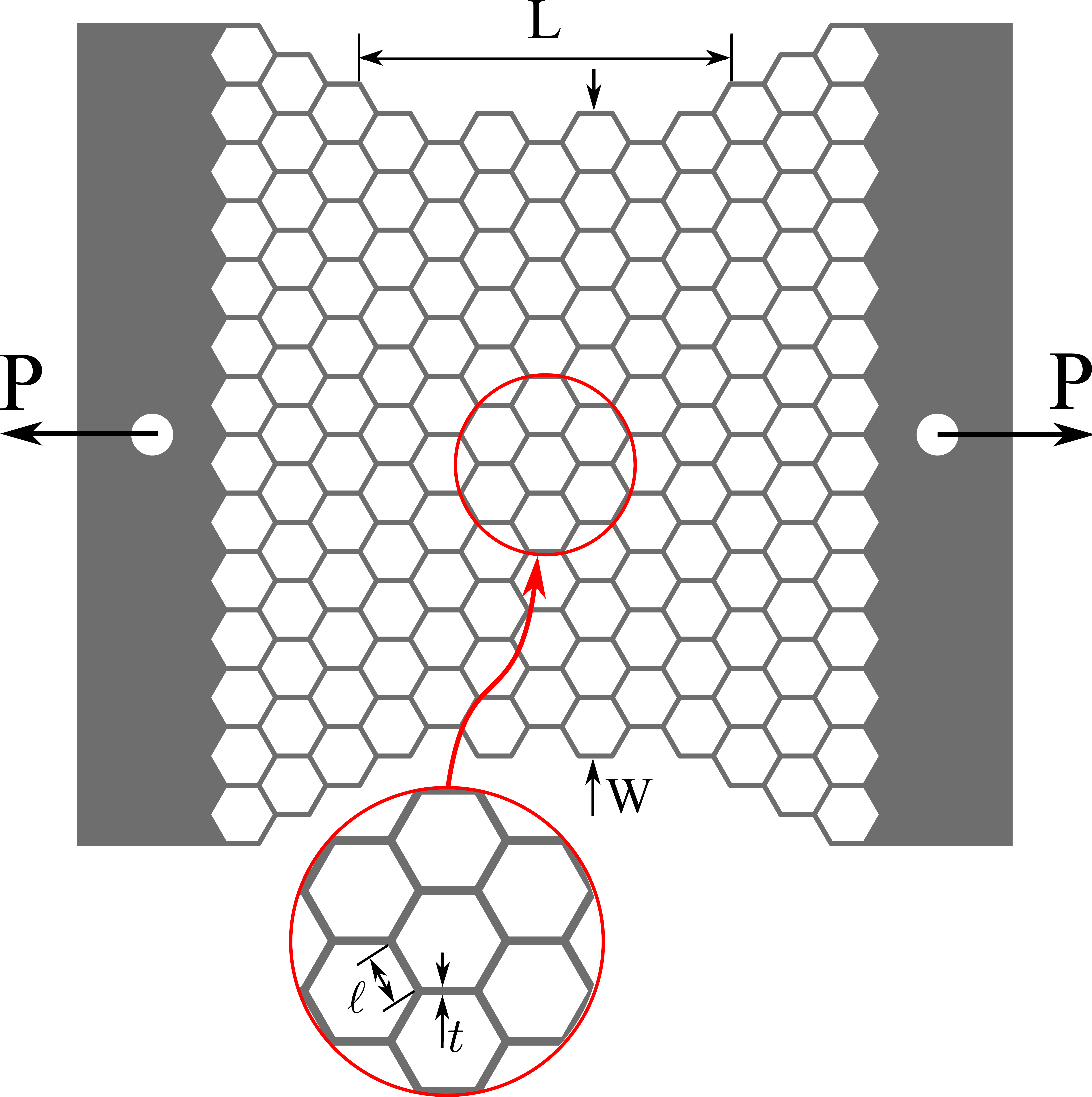}\label{fig:PerfectLatticeSketch}}
  \caption{\protect\subref*{fig:dbTension} Macroscale dogbone-shaped
    specimen, \protect\subref*{fig:singleStrutTension} single strut
    specimen, \protect\subref*{fig:singleStrut} unit cell test
    specimen, and \protect\subref*{fig:PerfectLatticeSketch} regular lattice specimen of $\bar\rho=0.11$. The sheet thickness of all samples is
    $B_0=\SI{5}{mm}$.}
  \label{fig:samples}
\end{figure}

\begin{figure}[htbp]
  \centering
  \sidesubfloat[]{\includegraphics[width=0.43\textwidth]{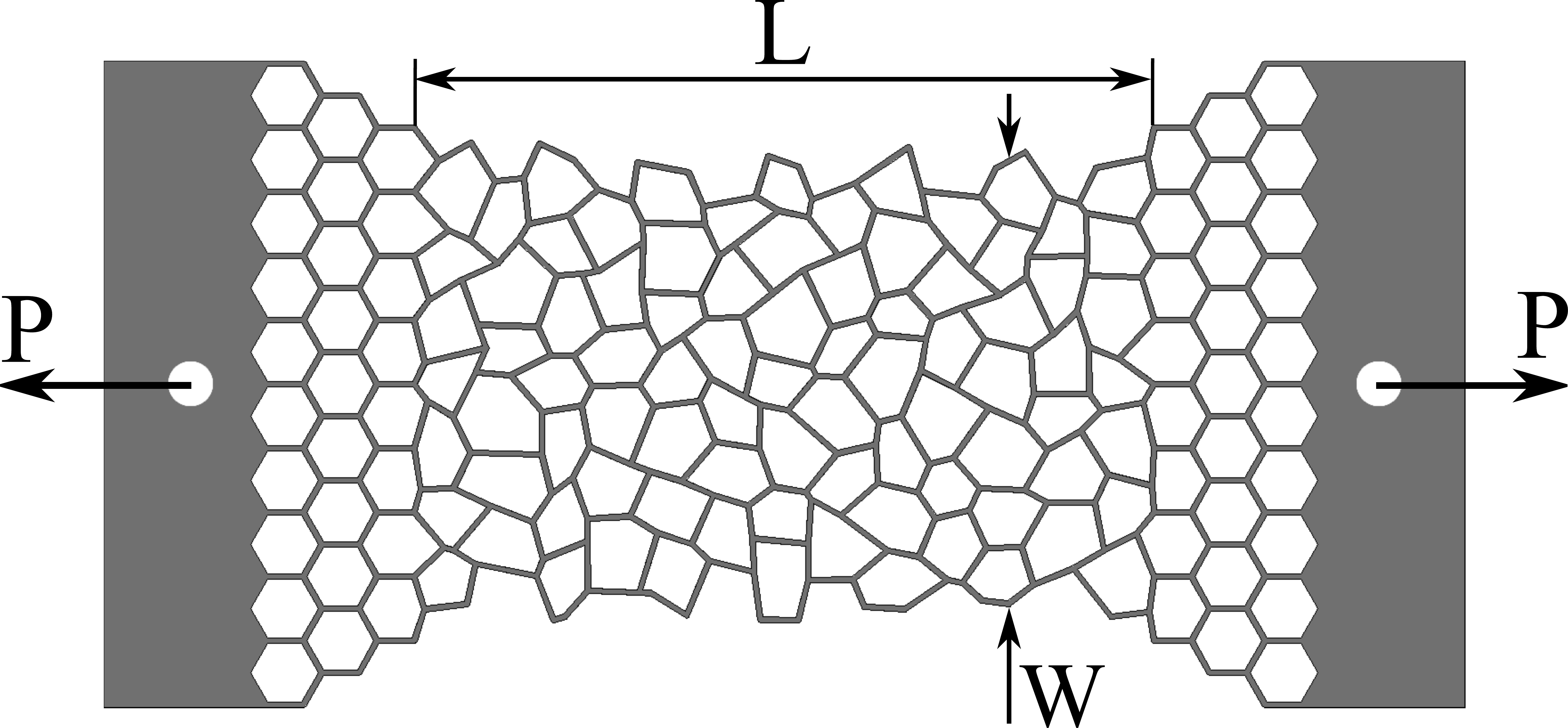}\label{fig:Model_Seed1_0_Radius0_5}}\hfill
  \sidesubfloat[]{\includegraphics[width=0.43\textwidth]{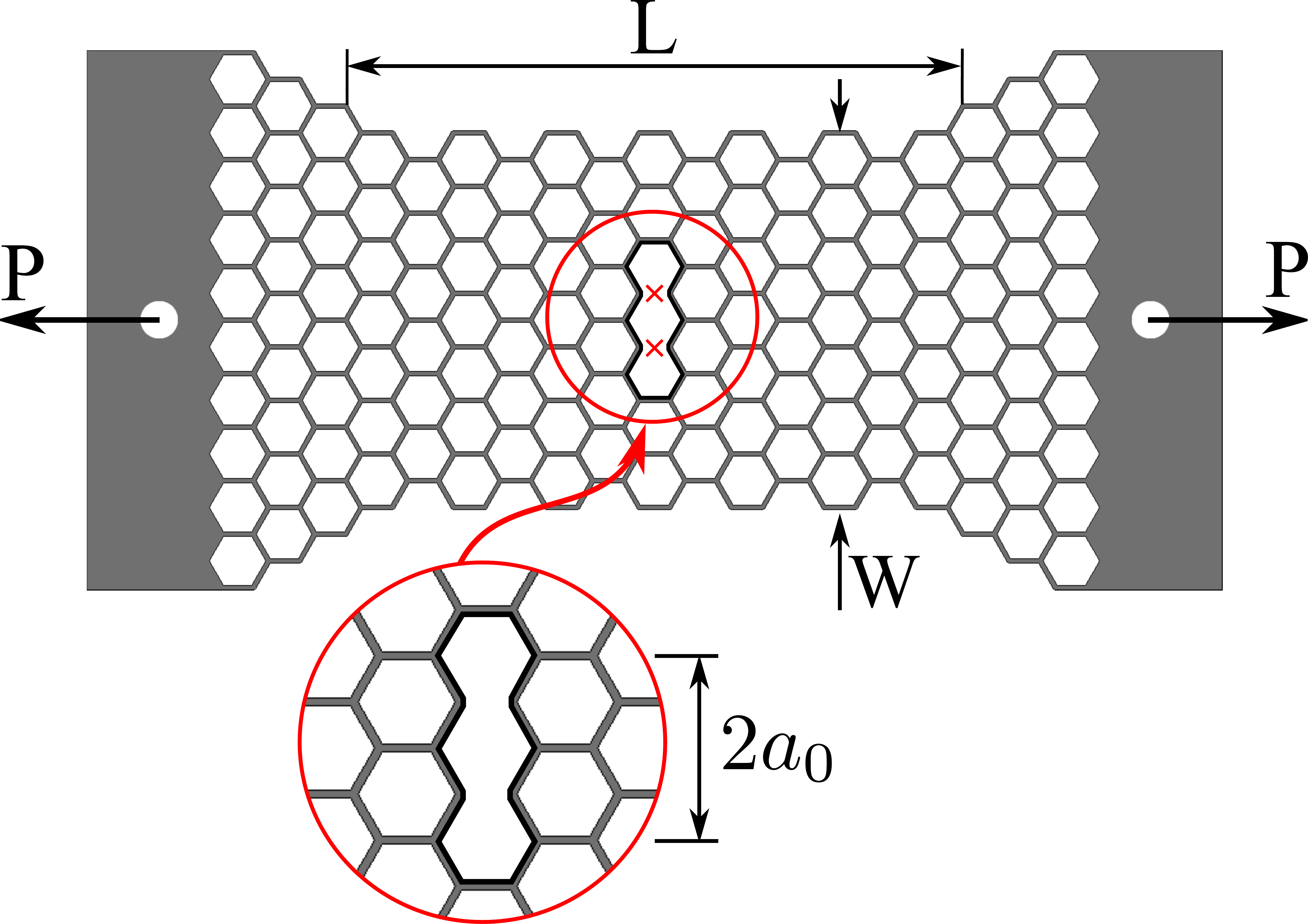}\label{fig:RemovedCellWallsSketch}}\\[\baselineskip]
  \sidesubfloat[]{\includegraphics[width=0.43\textwidth]{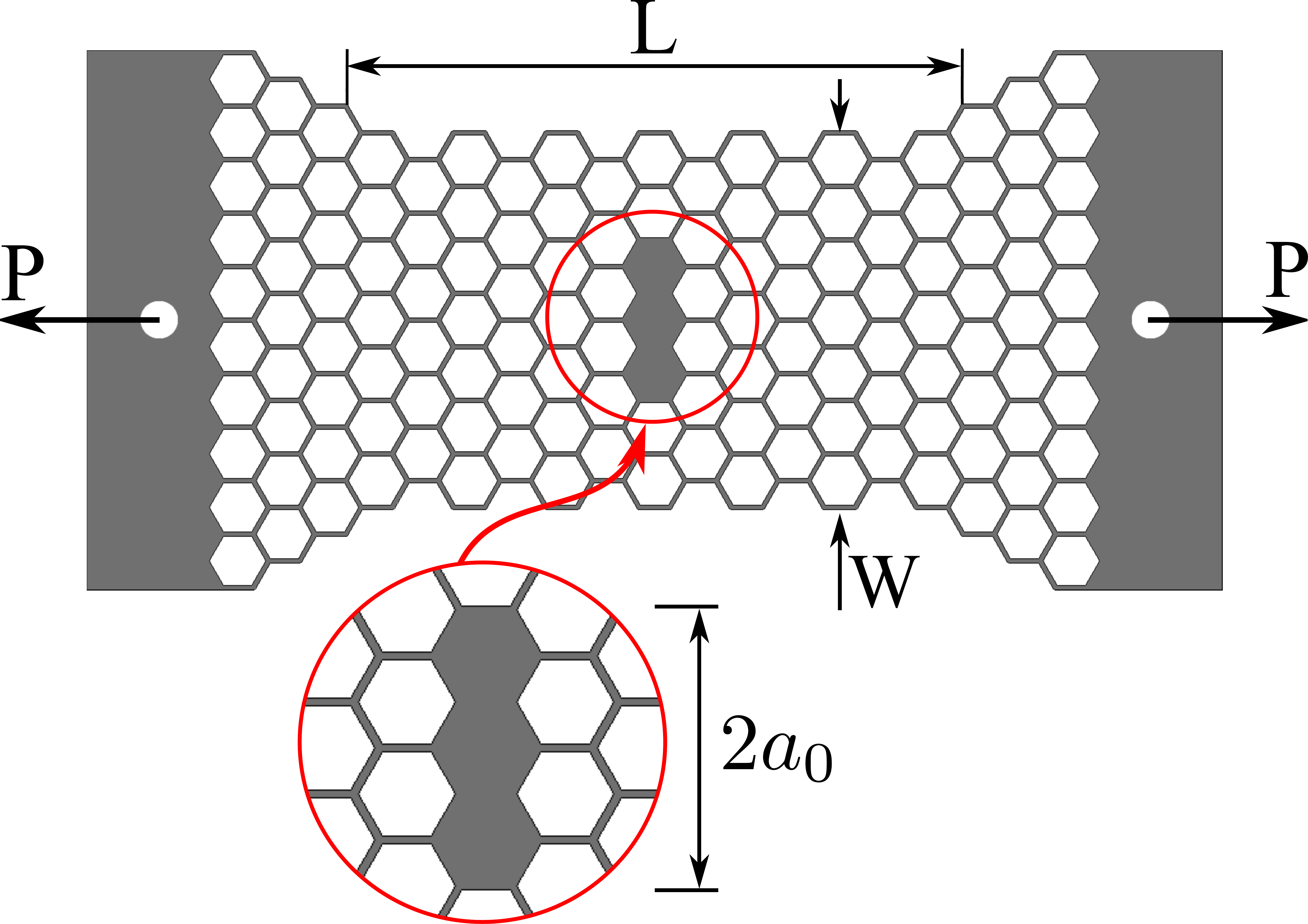}\label{fig:FilledCellsSketch}}
  \caption{Lattice specimens ($\bar\rho=0.11$, gauge width $W$ of 7
    cells, and gauge length $L$ of 13 cells) containing as-designed
    defects in the form of
    \protect\subref*{fig:Model_Seed1_0_Radius0_5} randomly perturbed
    joints ($R/\ell=0.5$),
    \protect\subref*{fig:RemovedCellWallsSketch} a row of missing cell
    walls, and \protect\subref*{fig:FilledCellsSketch} a row of solid
    inclusions.}
    \label{fig:geometries}
\end{figure}

\begin{figure}[htbp]
  \centering
  \includegraphics[width=0.55\textwidth]{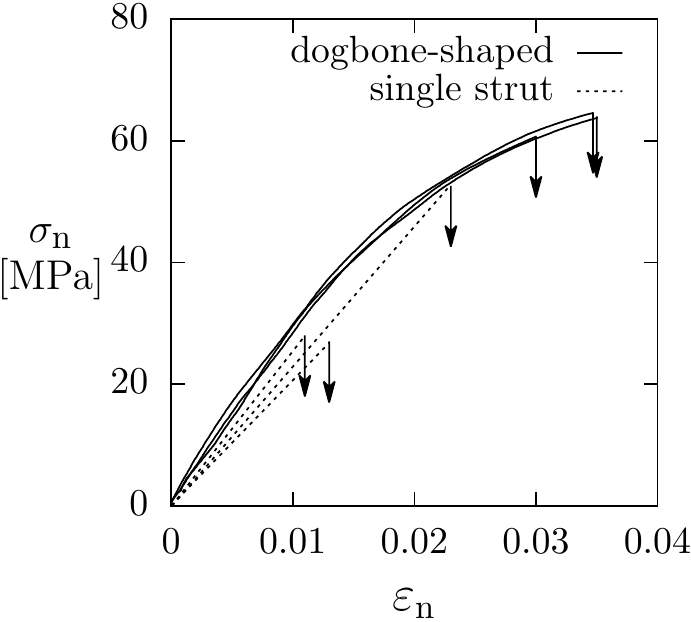}
  \caption{Nominal stress versus nominal strain response of  macroscale dogbone-shaped specimens and single 
    strut specimens.}
  \label{fig:stressStrain_PMMA_StrutVsSolid}
\end{figure}

\begin{figure}[htbp]
  \begin{center}
    \includegraphics[width=0.95\textwidth]{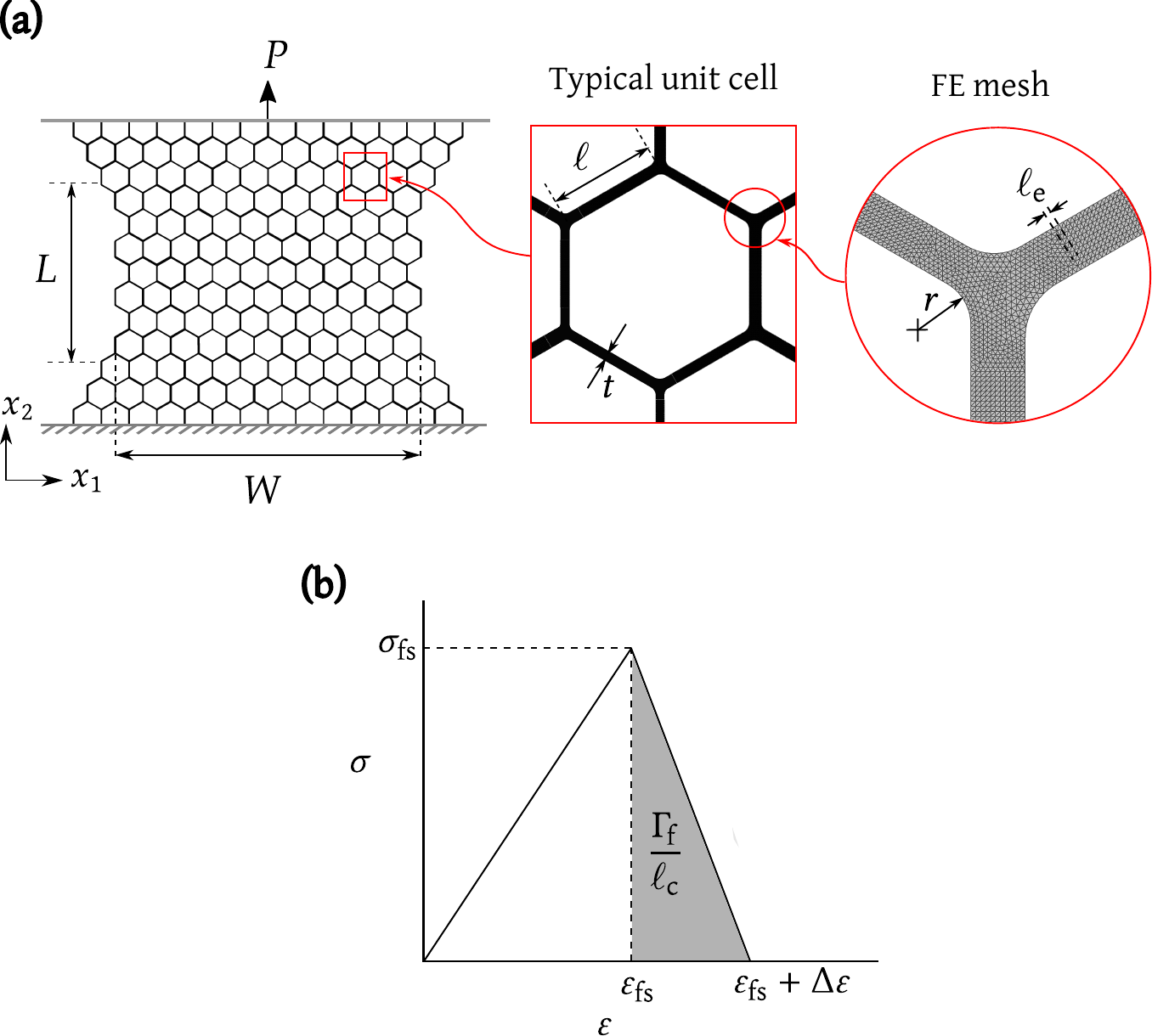}    
  \end{center}
  \caption{Details of the FE model: (a) geometry, loading, and
    boundary conditions employed in the FE simulations of lattice
    specimens under uniaxial tension. A typical unit cell within the
    lattice is shown along with the FE mesh for a joint. (b) Assumed
    stress versus strain response of the cell wall solid.}
  \label{fig:fe_sketch}
\end{figure}

\begin{figure}[htbp]
  \centering
  \sidesubfloat[]{\includegraphics[width=0.55\textwidth]{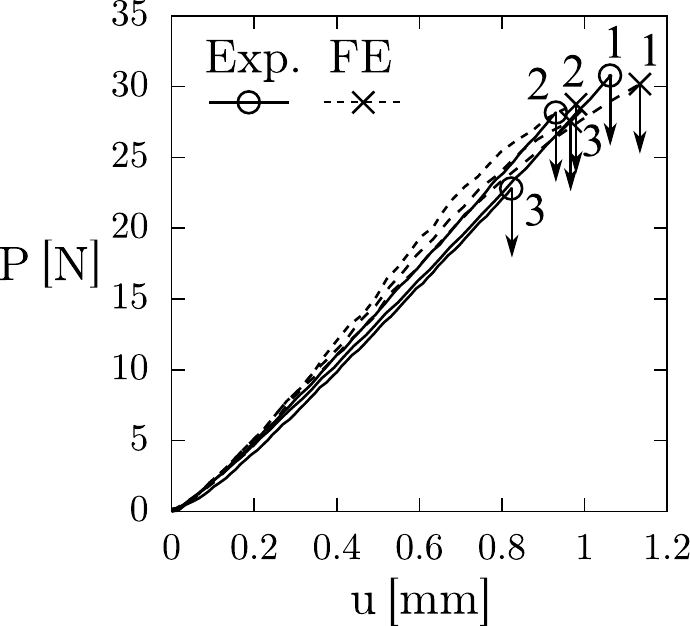}\label{fig:fe_pu_singlestrut}}\\[\baselineskip]
  \sidesubfloat[]{\includegraphics[width=0.65\textwidth]{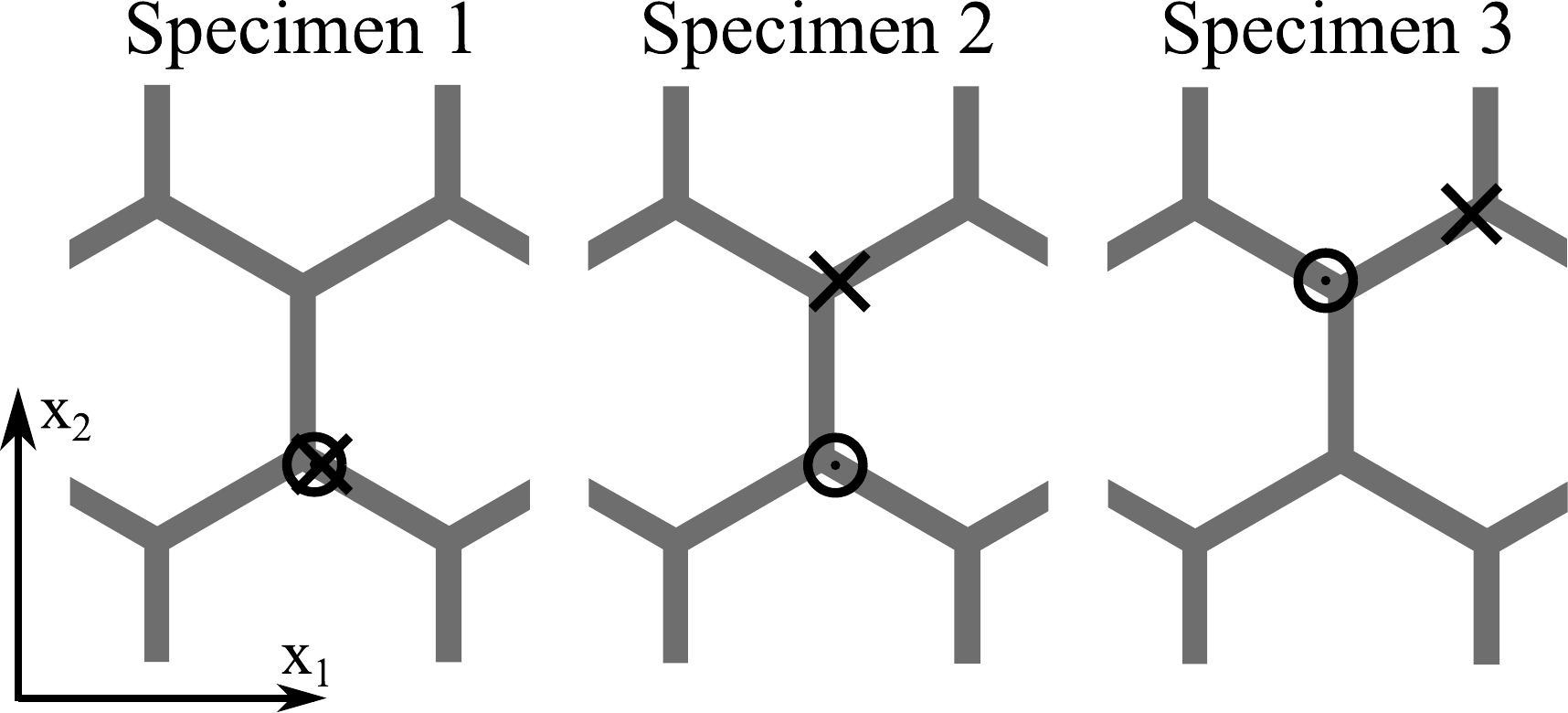}\label{fig:fe_failure_singlestrut}}
  \caption{Measured (Exp.) and predicted (FE) responses for 3
    as-manufactured unit cell specimens with $t/\ell=0.1$.
    \protect\subref*{fig:fe_pu_singlestrut} Load versus displacement
    response until first strut failure, and
    \protect\subref*{fig:fe_failure_singlestrut} location of first
    strut failure for the three specimens; the circles indicate
    observed failure sites while the crosses indicate predicted
    failure sites.}
  \label{fig:fe_singlestrut}
\end{figure}

\begin{figure}[htbp]
  \centering
  \sidesubfloat[]{\includegraphics[width=0.43\textwidth]{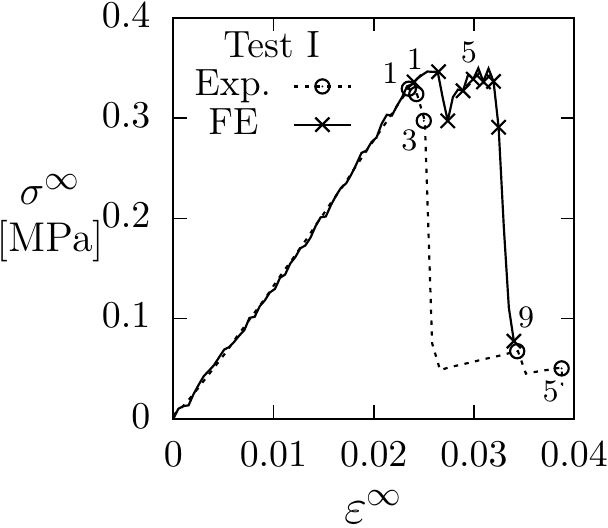}\label{fig:fe_exp_stressstrain_47}}\hfill
  \addtocounter{subfigure}{1}
  \sidesubfloat[]{\includegraphics[width=0.43\textwidth]{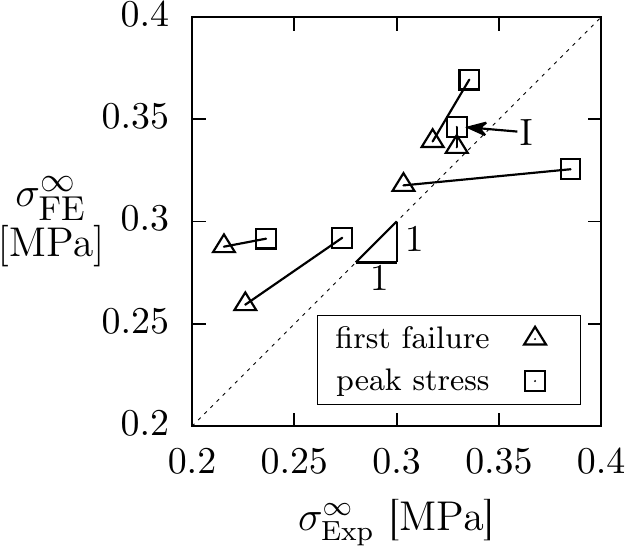}\label{fig:fe_exp_stresses_mod}}\\[\baselineskip]
    \addtocounter{subfigure}{-2}
  \sidesubfloat[]{\includegraphics[width=0.9\textwidth]{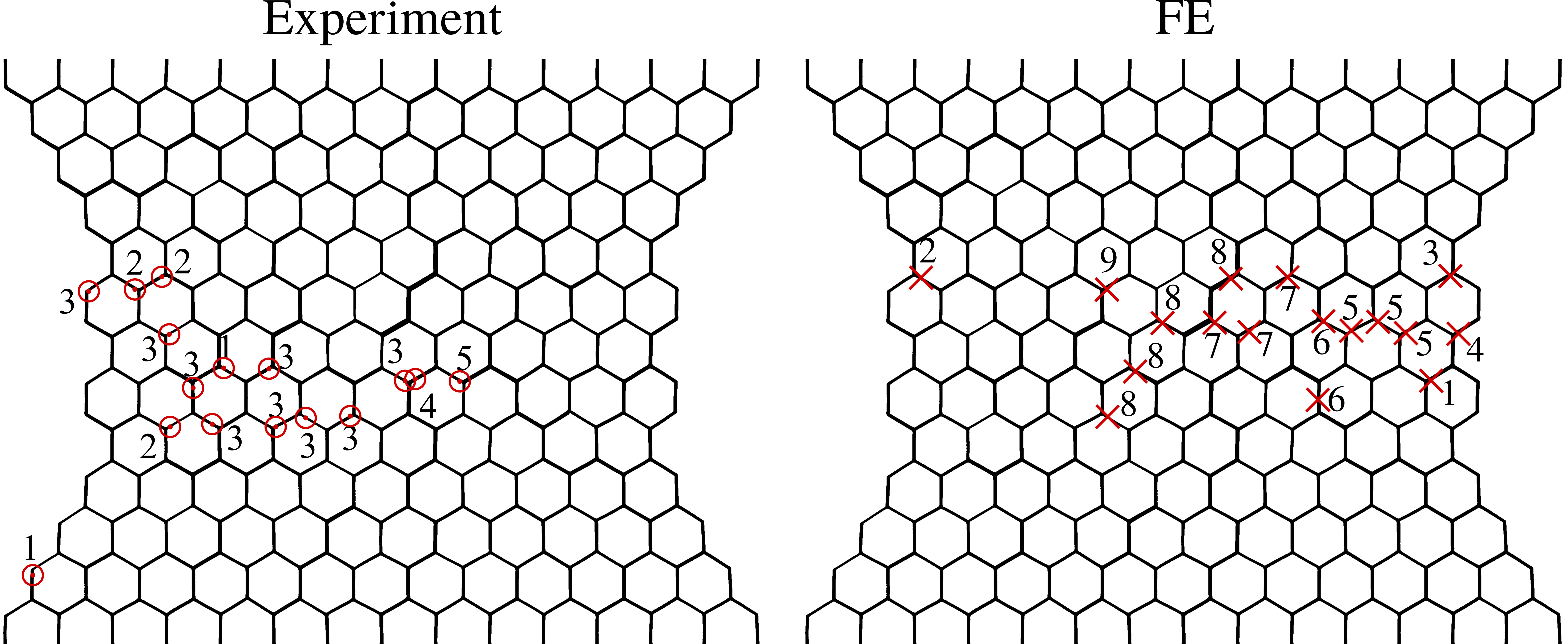}\label{fig:fe_failure_lattice_caseE}}\\[\baselineskip]
  \caption{Comparison of the measured and FE responses for lattice
    specimens of $\bar\rho=0.11$.
    \protect\subref*{fig:fe_exp_stressstrain_47} Stress versus strain
    response, and \protect\subref*{fig:fe_failure_lattice_caseE}
    sequence of strut failure for one realisation (Test~I). The
    crosses and circles indicate the location of failure and the
    number denotes its position in the failure sequence. The
    macroscopic stress at first strut failure and the peak stress of
    five measured specimens are compared with FE predictions in
    \protect\subref*{fig:fe_exp_stresses_mod} along with Test~I.}
  \label{fig:fe_lattice_caseE}
\end{figure}

\begin{figure}[htbp]
  \centering
  \sidesubfloat[]{\includegraphics[width=0.43\textwidth]{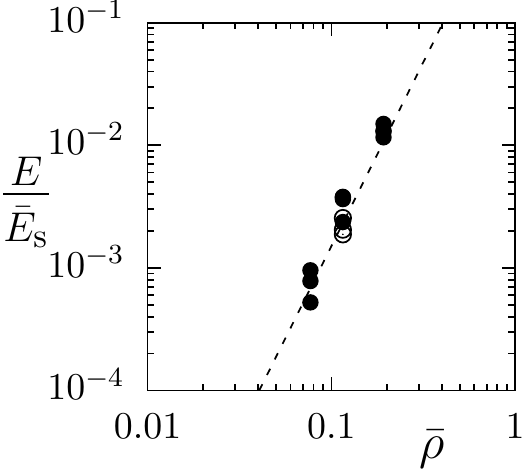}\label{fig:relativeDensity_YoungsModulus_perfectHex}}\hfill
  \sidesubfloat[]{\includegraphics[width=0.43\textwidth]{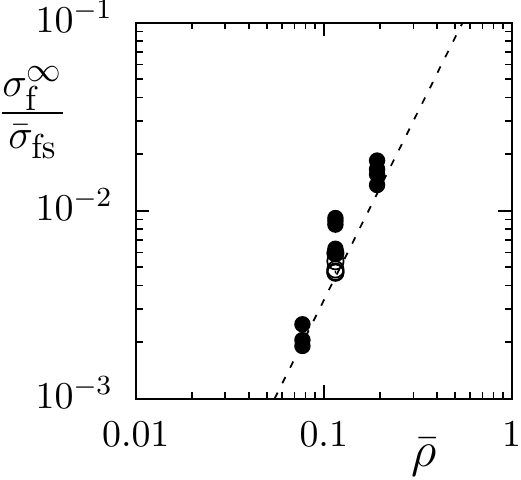}\label{fig:relativeDensity_stressFailure_perfectHex}}
  \caption{Measured values of
    \protect\subref*{fig:relativeDensity_YoungsModulus_perfectHex}
    macroscopic Young's modulus $E$, and
    \protect\subref*{fig:relativeDensity_stressFailure_perfectHex}
    macroscopic tensile strength $\sigma^\infty_\mathrm{f}$ of regular
    hexagonal lattice specimens as a function of
    $\bar{\rho}$. \changes{Unfilled symbols are additional
      measurements from results from regular lattice specimens of
      $7\,\times\,13$ cells and} dashed lines are predictions as given
    by \eqref{eq:youngs_modulus}.}
 \label{fig:scaling_laws}
\end{figure}

\begin{figure}[htbp]
  \centering
  \sidesubfloat[]{\includegraphics[width=0.44\textwidth]{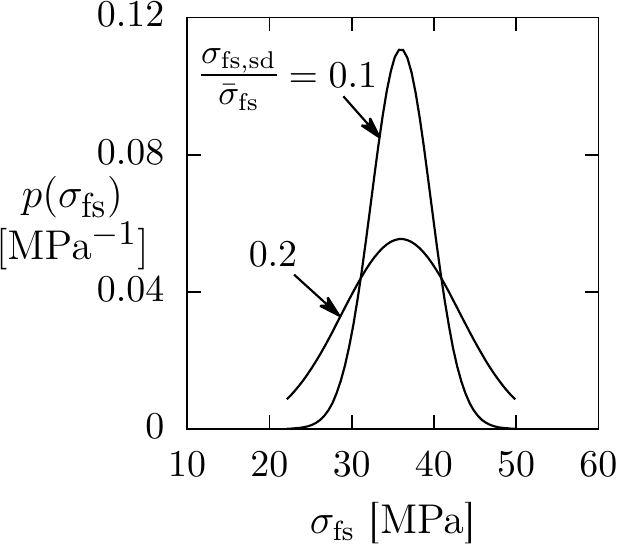}\label{fig:fe_pdf_strutdispersion}}\hfill
  \sidesubfloat[]{\includegraphics[width=0.4\textwidth]{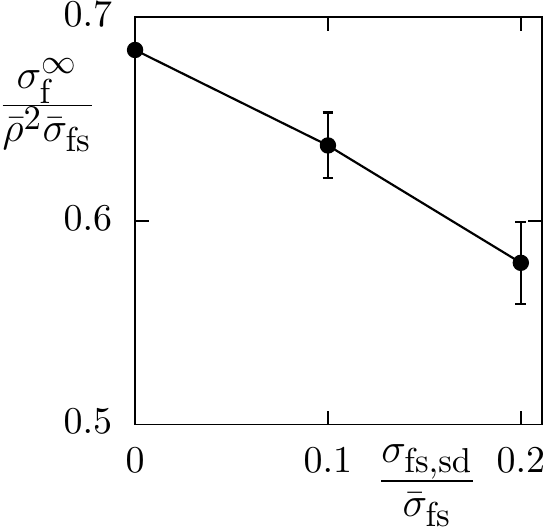}\label{fig:fe_sf_strutdispersion}}
  \caption{Role of dispersion in strut tensile strength within a
    regular lattice. \protect\subref*{fig:fe_pdf_strutdispersion}
    Assumed normal probability distribution functions for strut
    tensile strength $\sigma_{\rm fs}$, and
    \protect\subref*{fig:fe_sf_strutdispersion} the normalised
    macroscopic tensile strength as a function of the dispersion in
    strut strength $\sigma_\mathrm{fs,sd}/\bar{\sigma}_\mathrm{fs}$.}
  \label{fig:fe_strutdispersion}
\end{figure}

\begin{figure}[htbp]
  \centering
  \sidesubfloat[]{\includegraphics[width=0.55\textwidth]{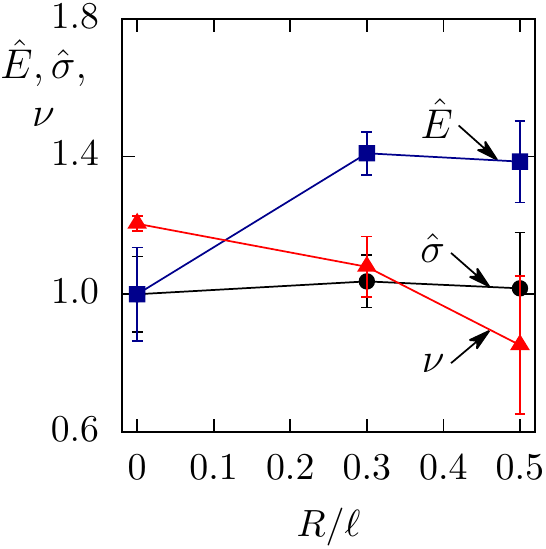}\label{fig:misplacedNodes_AllRTE}}
  \sidesubfloat[]{\includegraphics[width=0.35\textwidth]{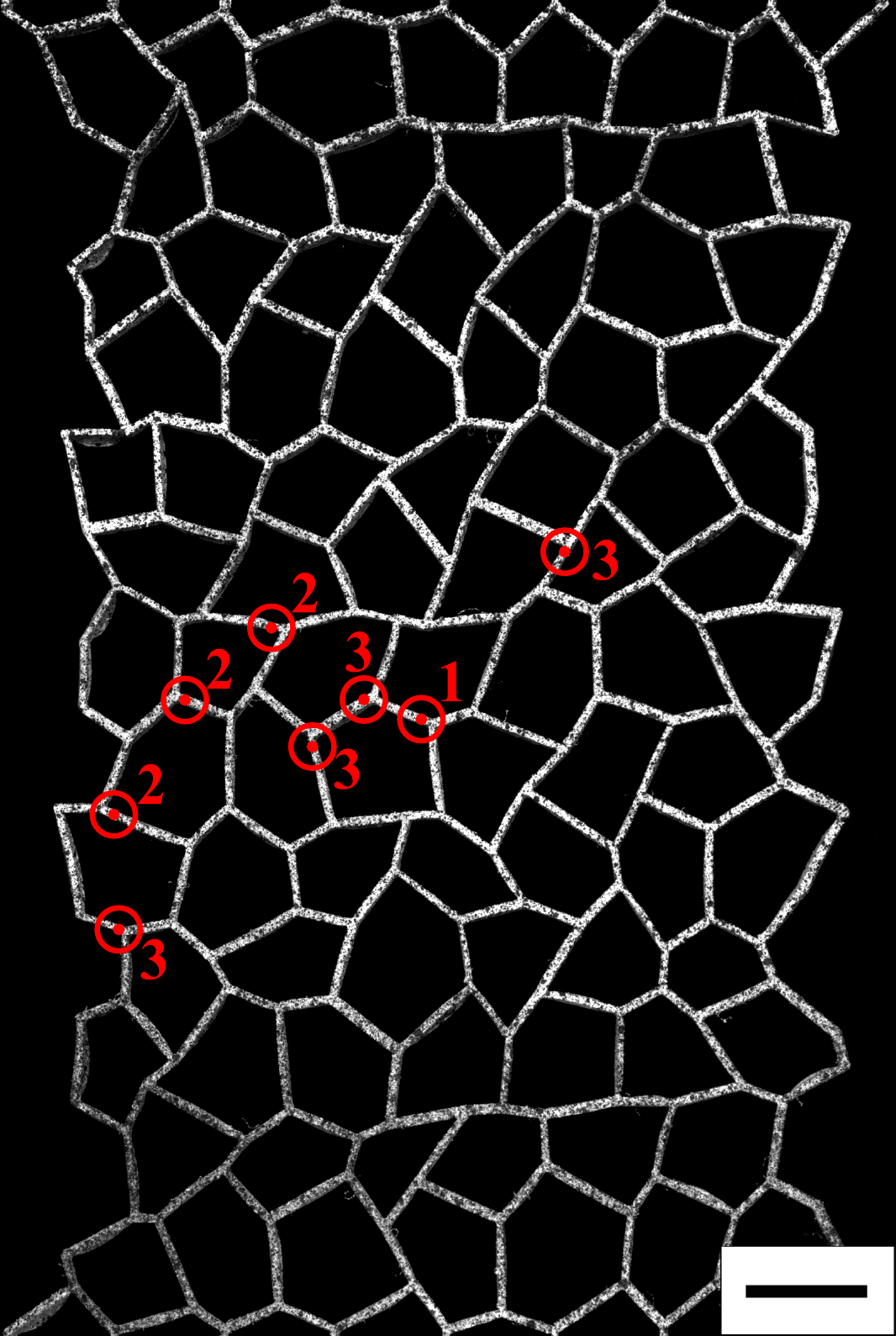}\label{fig:hexlattice_cut_26_failure}}
  \caption{\protect\subref*{fig:misplacedNodes_AllRTE} Measured
    macroscopic properties of imperfect lattices with randomly
    misplaced joints: normalised Young's modulus $\hat{E}$, normalised
    tensile strength $\hat{\sigma}$, and Poisson's ratio
    $\nu$. \protect\subref*{fig:hexlattice_cut_26_failure} Sequence of
    strut failure for one specimen of $R/\ell=0.5$. The scale bar is
    of length \SI{10}{mm}.}
  \label{fig:misplaced_nodes}
\end{figure}

\clearpage
\begin{figure}[htbp]
  \centering
  \sidesubfloat[]{\includegraphics[width=0.42\textwidth]{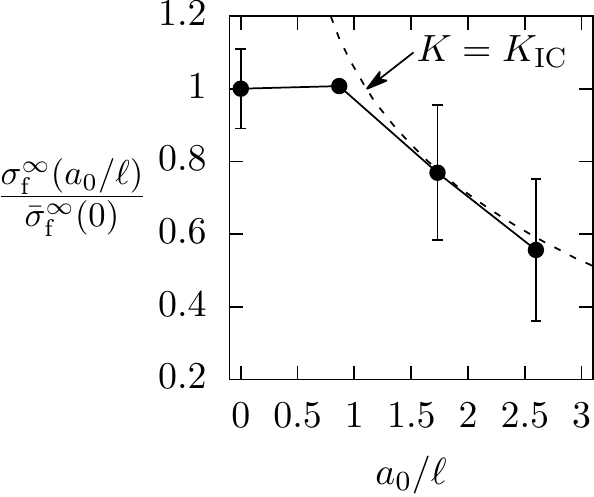}\label{fig:removed_cellswalls_stress}}\quad
  \sidesubfloat[]{\includegraphics[width=0.41\textwidth]{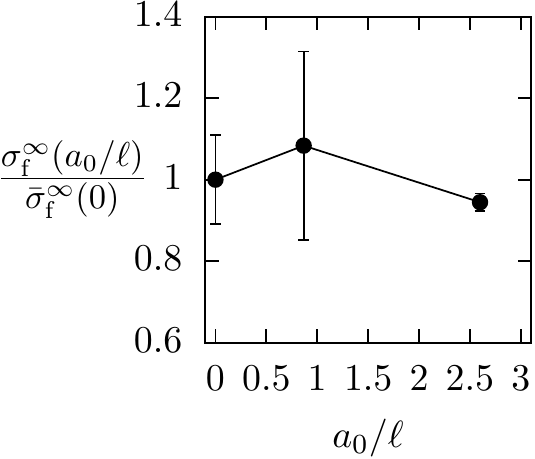}\label{fig:filled_cells_stress}}\\[\baselineskip]
  \sidesubfloat[]{\includegraphics[width=0.35\textwidth]{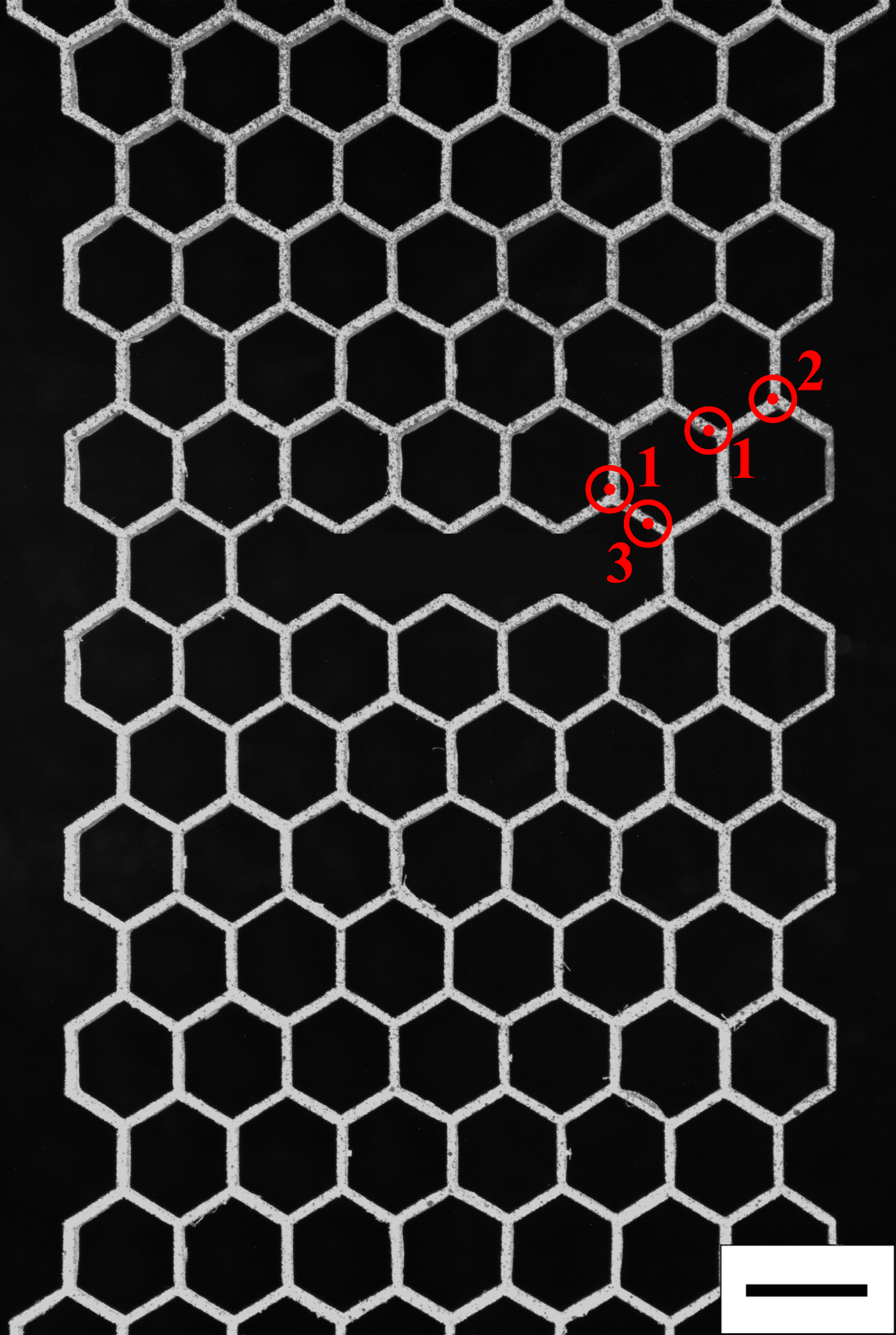}\label{fig:hexlattice_cut_32_failure}}\hfil
  \sidesubfloat[]{\includegraphics[width=0.35\textwidth]{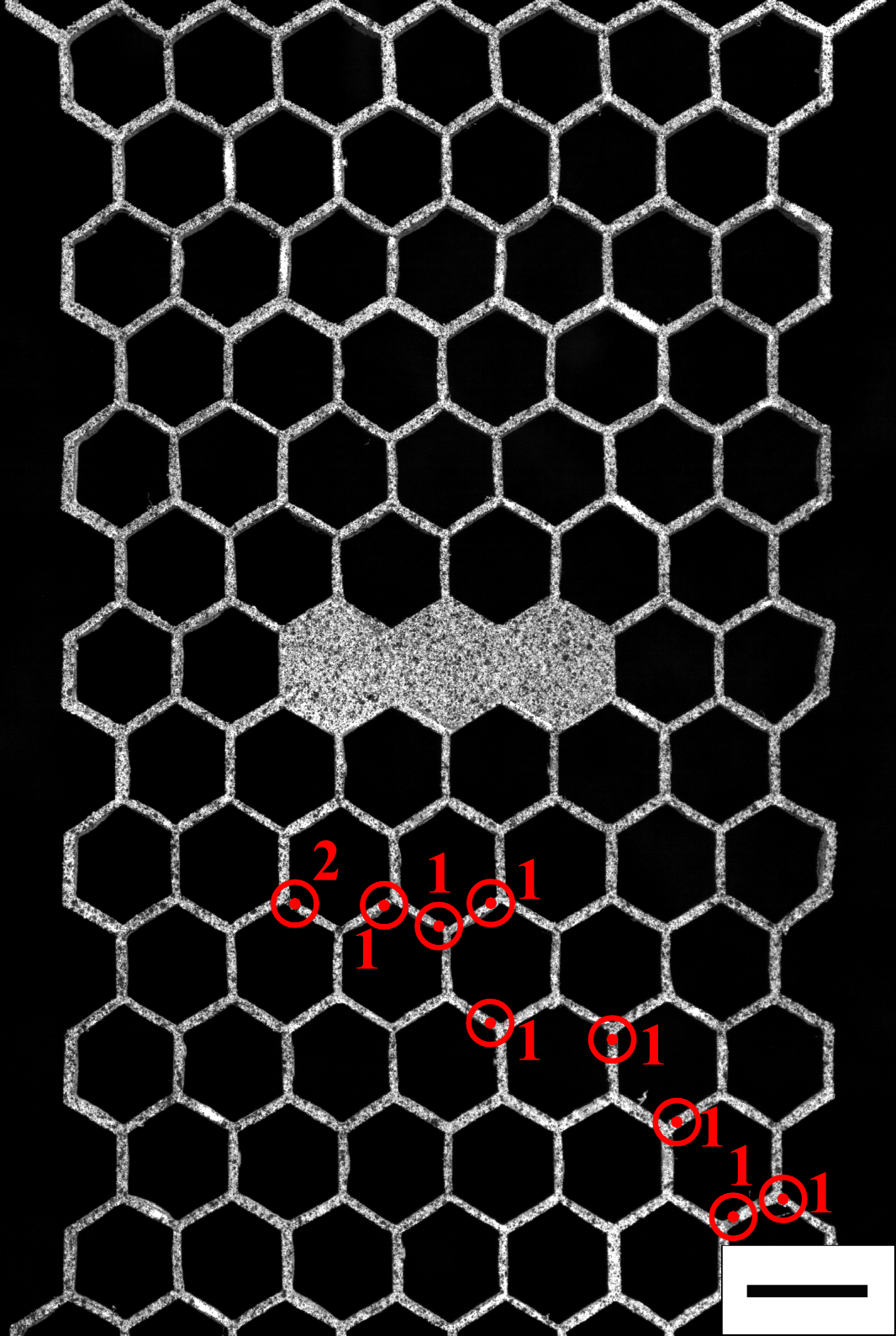}\label{fig:hexlattice_cut_21_failure}}
  \caption{Measured tensile strength of lattice specimens with
    \protect\subref*{fig:removed_cellswalls_stress} a row of missing cell walls
    and \protect\subref*{fig:filled_cells_stress} a row of solid
    inclusions, as a function of the defect size $a_{\rm
      0}/\ell$. Sequence of strut failure in one specimen containing
    \protect\subref*{fig:hexlattice_cut_32_failure} missing cell walls
    ($a_0/\ell=2.6$) or
    \protect\subref*{fig:hexlattice_cut_21_failure} solid inclusion
    ($a_0/\ell = 2.6$). The scale bar is of length \SI{10}{mm}.}
  \label{fig:imperfect_lattices}
\end{figure}
